\documentclass[12pt,preprint,psfig,epsf]{aastex}

\newcommand{\dmm}{\mbox{$\Delta$m$_{15}(B)$}}

\shorttitle{The Type~Ia Supernova 1999ac}
\shortauthors{Phillips et al.} 

\begin{document}
\received{17 January 2006}

\title{Optical and Near-Infrared Observations of the Peculiar
Type~Ia Supernova 1999ac\altaffilmark{1}}

\author{
Mark M. Phillips,\altaffilmark{2}
Kevin Krisciunas,\altaffilmark{3}
Nicholas B. Suntzeff,\altaffilmark{4}
R. G. Abraham,\altaffilmark{5}
M. G. Beckett,\altaffilmark{6}
Marco Bonati,\altaffilmark{4}
Pablo Candia,\altaffilmark{7}
T. Michael Corwin,\altaffilmark{8}
Darren L. Depoy,\altaffilmark{9}
Juan Espinoza,\altaffilmark{4}
Andrew E. Firth,\altaffilmark{10}
Wendy L. Freedman,\altaffilmark{5}
Gaspar Galaz,\altaffilmark{11}
Lisa Germany,\altaffilmark{12}
David Gonzalez,\altaffilmark{4}
Mario Hamuy,\altaffilmark{14}
N. C. Hastings,\altaffilmark{13}
Aimee L. Hungerford,\altaffilmark{15}
Valentin D. Ivanov,\altaffilmark{12}
Erika Labb\'{e},\altaffilmark{11}
Ronald O. Marzke,\altaffilmark{6}
Patrick J. McCarthy,\altaffilmark{5}
Richard G. McMahon,\altaffilmark{10}
Russet McMillan,\altaffilmark{13}
Cesar Muena,\altaffilmark{2}
S. E. Persson,\altaffilmark{5}
Miguel Roth,\altaffilmark{2}
Mar\'{i}a Teresa Ruiz,\altaffilmark{14}
R. Chris Smith,\altaffilmark{4}
Roger Smith,\altaffilmark{16}
Louis-Gregory Strolger,\altaffilmark{17}
and Christopher Stubbs\altaffilmark{18}
}
\altaffiltext{1}{Based in part on observations taken at the Cerro Tololo
Inter-American Observatory, National Optical Astronomy Observatory, 
which is operated by the Association of Universities for Research in 
Astronomy, Inc. (AURA) under cooperative agreement with the National 
Science Foundation.
This paper is also based in part on observations obtained with the Apache
Point Observatory 3.5-meter telescope, which is owned and operated
by the Astrophysical Research Consortium.}
\altaffiltext{2}{Las Campanas Observatory, Carnegie Observatories, 
  Casilla 601, La Serena, Chile; {mmp@lco.cl}, {miguel@lco.cl} }
\altaffiltext{3}{University of Notre Dame, Department of Physics, 225
  Nieuwland Science Hall, Notre Dame, IN 46556-5670;
  {kkrisciu@nd.edu} }
\altaffiltext{4}{Cerro Tololo Inter-American Observatory, Casilla 603,
  La Serena, Chile; {nsuntzeff@noao.edu}, {jespinoza@ctio.noao.edu}, 
  {dgonzalez@ctio.noao.edu}, {csmith@ctio.noao.edu} }
\altaffiltext{5}{Observatories of the Carnegie Institution of Washington, 813 Santa 
  Barbara Street, Pasadena, CA 91101; {wendy@ociw.edu}, {pmc2@ociw.edu},
  {persson@ociw.edu} }
\altaffiltext{6}{San Franciso State University, Physics and Astronomy, 1600 
  Holloway Avenue, San Francisco, CA 94132-4163; {marzke@stars.sfsu.edu} }
\altaffiltext{7}{Gemini Operations Center, Casilla 603, La Serena, Chile; 
  {pcandia@gemini.edu}}
\altaffiltext{8}{University of North Carolina at Charlotte, Department of Physics, 
  Charlotte, NC 28223; {mcorwin@uncc.edu} }
\altaffiltext{9}{Ohio State University, Department of Astronomy, 140 W. 18th 
  Avenue,  Columbus, OH 43210; {depoy@astronomy.ohio-state.edu} }
\altaffiltext{10}{Institute of Astronomy, Madingley Road, Cambridge, CB3 0HA, 
  England, UK; {rgm@ast.cam.ac.uk} }
\altaffiltext{11}{Departamento de Astronom\'{i}a y Astrofisica,
  Pontificia Universidad Catolica de Chile, Casilla 306, Santiago 22, Chile; 
  {ggalaz@astro.puc.cl}, {elabbe@astro.puc.cl} }
\altaffiltext{12}{European Southern Observatory, Avenida 
  Alonso de Cordova 3107, Vitacura, Casilla 19001, Santiago 19, Chile;
  {lgermany@eso.org}, {vivanov@eso.org} }
\altaffiltext{13}{Apache Point Observatory, P. O. Box 59, Sunspot, NM 88349-0059;
  {hastings@apo.nmsu.edu}, {rmcmillan@apo.nmsu.edu} }
\altaffiltext{14}{Universidad de Chile, Departmento de Astronom\'{i}a, 
  Casilla 36-D, Santiago, Chile; {mhamuy@das.uchile.cl}, {mtruiz@das.uchile.cl} }
\altaffiltext{15}{CCS-4, MS D409, Los Alamos National Laboratory, Los Alamos, NM 
  87545; {aimee@lanl.gov} }
\altaffiltext{16}{California Institute of Technology, Astronomy Department,
  MS 105-24, Pasadena, CA 91125; {rsmith@astro.caltech.edu} }
\altaffiltext{17}{Department of Physics and Astronomy, Western Kentucky University,
  1906 College Heights Blvd. \#11077, Bowling Green, KY 42101-1077; {louis.strolger@wku.edu} }
\altaffiltext{18}{Department of Physics and Department of Astronomy,
  17 Oxford Street, Harvard University, Cambridge MA 02138; {cstubbs@fas.harvard.edu}}

\begin{abstract} 
We present 39 nights of optical photometry, 34 nights of infrared photometry,
and 4 nights of optical spectroscopy of the Type Ia SN 1999ac.  This supernova
was discovered two weeks before maximum light, and observations were begun
shortly thereafter.  At early times its spectra resembled the unusual
SN~1999aa and were characterized by very high velocities in the \ion{Ca}{2}
H\&K lines, but very low velocities in the \ion{Si}{2} $\lambda$ 6355 line.  
The optical photometry showed a slow rise to peak brightness but, quite
peculiarly, was followed by a more rapid decline from maximum. Thus,
the $B$- and $V$-band light curves cannot be characterized by a single stretch
factor.  We argue that the best measure of the nature of this object is {\em
not} the decline rate parameter \dmm.  The $B-V$ colors were unusual from 30
to 90 days after maximum light in that they evolved to bluer values at a much
slower rate than normal Type Ia supernovae.  The spectra and bolometric light
curve indicate that this event was similar to the spectroscopically peculiar
slow decliner SN~1999aa.
\end{abstract}


\section{Introduction}

Due to their considerable potential as extragalactic standard candles, Type Ia
supernovae (SNe~Ia) have been the subject of intense study during the last 15
years.  Although it has by now been amply demonstrated that SNe~Ia cover a
range in instrinsic brightness of a factor of two or 
more \citep[see][and references therein] {Lei00},
the fortuitous existence of a correlation
between the peak luminosity and the rate of decline from maximum of the $B$
light curve, including prescriptions for color corrections \citep{Phi93,
Ham_etal96b, Rie_etal96, Phi_etal99, Nob_etal03, Guy_etal05}, 
has allowed distances to be measured to
better than 10 percent to redshifts $z \leq 0.1$ \citep{Ham_etal96a,
Rie_etal96}. Because they can be observed to very large distances, SNe~Ia have
become the tool of choice for measuring the Hubble constant
\citep[e.g.,][]{Ham_etal96a, Rie_etal96, San_etal96, Jha_etal99, Sun_etal99,
Phi_etal99, Tri_Bra99, Gib_etal00, Par_etal00, Fre_etal01, Phi_etal03,
Rie_etal05} and, in combination with measurements of fluctuations in the
microwave background radiation \citep{Ben_etal03}, have provided compelling
evidence for an accelerating universe \citep{Gar_etal98, Rie_etal98,
Per_etal99, Ton_etal03, Kno_etal03, Bar_etal04, Rie_etal04, Kri_etal05}.

Although the majority of SNe~Ia display very similar spectral evolution, 
roughly one-third of all events show spectroscopic
peculiarities \citep{Li_etal01a}.  Three well-observed examples -- SN~1991T,
SN~1991bg, and SN~1986G -- have served as the prototypes of
spectroscopically-peculiar SNe~Ia \citep{Bra_etal93}.

The pre-maximum optical spectra of SN~1991T showed unusually weak lines of
\ion{Si}{2}, \ion{S}{2}, and \ion{Ca}{2}, while at the same time displaying
prominent high-excitation features of \ion{Fe}{3} \citep{Fil_etal92a,
Phi_etal92}. The \ion{Si}{2}, \ion{S}{2}, and \ion{Ca}{2} lines grew quickly in
strength following maximum until the spectrum appeared essentially normal a few
weeks past maximum.  SN~1991T dimmed slowly from maximum and was originally
thought to be considerably more luminous than ``normal'' SNe~Ia
\citep{Fil_etal92a, Phi_etal92}, although the Cepheid distance to the host
galaxy, NGC~4527, obtained later with the Hubble Space Telescope, implies that
this supernova was only moderately over-luminous \citep{Sah_etal01, 
Gib_Ste01}.

The main spectroscopic peculiarity of SN~1991bg was the presence at maximum light
of a broad absorption trough at 4100-4400~\AA~due mostly to \ion{Ti}{2} lines
\citep{Fil_etal92b, Lei_etal93, Tur_etal96, Maz_etal97}, reflecting a lower
effective temperature. As a consequence,
SN~1991bg was unusually red ($B-V \simeq 0.75$) at maximum light.  SN~1986G was
similar to 1991bg, but less extreme \citep{Phi_etal87, Cri_etal92}. Both events
were substantially sub-luminous.

\citet{Nug_etal95} have shown that the photometric sequence of SNe~Ia (i.e., the
luminosity vs. decline rate relation) also manifests itself as a spectroscopic
sequence which can be modelled in terms of a range in the effective
temperature at maximum.  The luminous, slowly-declining SN~1991T represents the
high-temperature extreme of the sequence, while the fast-declining, sub-luminous
SN~1986G and SN~1991bg correspond to the low-temperature limit.  Under this
scenario, none of these three SNe~Ia should be considered ``peculiar''.
However, not all luminous, slow-declining SNe~Ia display SN~1991T-like
pre-maximum spectra -- e.g., SN~1992bc \citep{Maz_etal94} and SN~1999ee
\citep{Ham_etal02} -- so it is not clear
if this interpretation is fully correct.  Moreover, the last few years have
revealed the existence of several SNe~Ia which are of an intermediate type
between SN~1991T events and ``normal'' SNe~Ia.  The prototype of these objects,
which may be more common than SN~1991T-like events, is SN~1999aa
\citep{Kri_etal00, Li_etal01a, Gar_etal04}. SN~1999aa was similar 
to SN~1991T in displaying
very weak \ion{Si}{2} 6150~\AA~absorption and prominent \ion{Fe}{3} lines in its
pre-maximum spectra, but it showed strong \ion{Ca}{2} H~\&~K absorption at the
same epoch in which this feature was weak or absent in SN~1991T
\citep{Fil_etal99}.  By maximum light the spectrum of SN~1999aa had evolved to
that of a ``normal'' Type~Ia supernova.  Hence, SN~1999aa-like events would be
difficult, if not impossible, to distinguish spectroscopically unless a spectrum
were obtained at least a week before maximum.

In order to more fully understand the range of spectroscopic and photometric
characteristics of SNe~Ia, we have organized a number of observing campaigns to obtain
optical photometry, infrared photometry, and optical spectroscopy of nearby events ($z
\leq 0.05$).  In this paper, we present observations obtained of the SN~1999aa-like
object SN~1999ac. SN~1999ac was discovered in the Sc galaxy NGC~6063 by
\citet{Mod_etal99} from images obtained on 1999 February 26.5 and 27.5 UT.  
It was
located at RA = 16:07:15.0, DEC = +07:58:20 (equinox 2000), some 24 arcsec east and 30
arcsec south of the core of its host. SN~1999ac was confirmed to be a Type~Ia
supernova by \citet{Phi_Kun99} from a spectrum taken on February 28 UT, who also noted
that the spectrum was similar to that of the slow decliner SN~1991T
\citep{Lir_etal98}. Like SN~1991T, the \ion{Fe}{3} lines at 4300 and 5000~\AA~in the
spectrum of SN~1999ac were observed to be strong and the \ion{Si}{2}
6355~\AA~absorption weak; but unlike 1991T, the \ion{Ca}{2} H~\&~K absorption in
SN~1999ac was strong and well-developed.  Phillips \& Kunkel speculated that SN~1999ac
had been caught near, or a few days before, maximum; our observations show that $B$
maximum did not actually occur until 13 days after this first spectrum was
obtained.

In \S2 we present our optical and near-infrared photometry of SN~1999ac.
The resulting light curves, which are among the most complete ever obtained
of a SN~Ia for the first few months following explosion, are contrasted
with the light curves of other well-observed SNe~Ia.
In \S3 we discuss the optical spectra obtained, likewise comparing these
with similar observations of both spectroscopically ``normal'' 
and ``peculiar'' SNe~Ia.  In \S4 we derive the most likely
host galaxy reddening for SN~1999ac, and compare the peak luminosity and 
bolometric light curve with those of more typical SNe~Ia.  
%In this same section, we attempt to understand the physical relationship 
%of SN~1999ac to the general family of SNe~Ia events.
Finally, in \S5 the conclusions of this 
investigation are summarized.

\section{Optical and Infrared Photometry}

\subsection{Observations and Data Reduction}

Optical $BV(RI)_{KC}$ imaging of SN~1999ac commenced at the Cerro
Tololo Inter-American Observatory (CTIO) on 1999 March 1 UT, and observations
were continued for nearly five months.\footnote[19]{Our $R$- and $I$-band 
data were obtained with ``Kron-Cousins'' filters \citep{Bes79}.  For the
sake of simplicity we shall drop the $KC$ subscripts when referring to
$RI$ data.}  Three different telescopes -- the
0.9~m, 1.5~m, and the Yale-AURA-Lisbon-Ohio
(YALO)~1.0~m -- were employed.  In addition, three nights
of optical imaging were obtained with the Apache Point Obervatory (APO) 
3.5~m telescope.

In Fig. \ref{finder} we show an optical finding chart of the field.  A sequence of
local ``standards'' in the supernova field was established from observations
obtained on multiple photometric nights.  The $BV(RI)_{KC}$ magnitudes for these
stars are given in Table \ref{standards} and are tied to observations of
\citet{Lan92} standards obtained on the same nights.  We reduced the optical data
for SN~1999ac with respect to the local standards using color coefficients derived
from the observations of the Landolt stars.  The local standards were verified to
be constant over the $\sim$5 month duration of our observation. This method of
data reduction allows one to observe the SN even under non-photometric conditions,
under the explicit assumption that terrestrial clouds are grey (i.e. do not
selectively filter out light of one color more than another).  In our experience
this is demonstrably valid as long as the exposures are not extremely short, or
the clouds extremely thick.  For more on the greyness of terrestrial clouds see
\citet{Ser70}, \citet{Wal_etal71}, and \citet{Ols83}.

We present $BV(RI)_{KC}$ photometry of SN~1999ac in Table \ref{opt_photom}.
All magnitudes were determined with {\sc daophot} \citep{Ste87, Ste90} using
PSF (point spread function) photometry on direct, non-subtracted images.
The brightness and location of the SN did not require the use of host galaxy
subtraction templates.

Near-infrared imaging of SN~1999ac in the $J_sHK_s$ bands was obtained over a
three month period beginning on the night of 1999 February 28 UT.  Most of the
data were obtained at the Las Campanas Observatory (LCO) with the Swope 1.0~m
and du~Pont 2.5~m telescopes.  A few images in $H$ and $K_s$ were also
obtained with the Steward Observatory (SO) 2.3~m and 1.5~m telescopes.  In a
similar manner to the optical photometry, a sequence of local standards was
established from observations made with the LCO 1.0~m telescope on photometric
nights. In Table \ref{ir_stds} we give the near infrared magnitudes for the
seven field stars, which are tied to the infrared standards of
\citet{Per_etal98}.  Two of these stars (IR3 and IR7) were very red and were
too faint in the optical bands to be useful as optical standards, but IR3 was
the principal star used to reduce the infrared photometry.  From a comparison
of the differential magnitudes of IR3 with respect to field stars 1 and 8, we
established that IR3 was constant at the 0.03 mag level during the period of 
our observations.

The final $J_sHK_s$ magnitudes of SN~1999ac are given in Table \ref{ir_photom}.
These were derived using the same PSF photometry software employed for the optical
data.  Since most of the IR imaging was obtained with the LCO 1.0~m telescope and
infrared camera, which is the very telescope and instrument used to establish the
system of standards of \citet{Per_etal98}, no color corrections were applied to
the IR photometry.  We also note that the system of \citet{Per_etal98} uses
different filters than the standard J-band and K-band filters.  The J$_s$ and
K$_s$ filters (where ``s'' stands for ``short'') have slightly different effective
wavelengths and narrower bandwidths than the corresponding Johnson filters.

\subsection{Optical Photometry}

The $BVRI$ light curves of SN~1999ac are shown in Fig. \ref{bvri}.
The symbols identify the data taken with the different telescopes
employed.  Also included are the published observations of \citet{Jha02}.

It may be seen that our optical photometry began $\sim$12 days before $B$
maximum was reached, providing an uninterrupted record of the rise to peak
brightness in each band and continuing without significant gaps to $\sim$4
months after $B$ maximum.  Except for the YALO 1.0~m observations, the data
obtained with the various telescopes, including the published measurements of
\citet{Jha02}, are in generally good agreement.  The YALO 1.0~m photometry
shows significant discrepancies, most notably in the $R$ band, but also
clearly present in $B$, $V$, and $I$, particularly near maximum light.  These
differences are accentuated in Fig. \ref{colors} which shows the evolution of
the $B-V$, $V-R$, and $V-I$ colors.  Note that the early $B-V$ points from the
YALO 1.0~m are nearly 0.2~mag redder than the trend defined by the 0.9~m and
\citet{Jha02} photometry, but that this difference has essentially disappeared
by JD~2,451,295.  In $V-R$ the differences are even larger and persist during
the entire period that the YALO 1.0~m was utilized.  In $V-I$ the agreement is
the best, but this is fortuitous since the YALO 1.0~m $V$ and $I$ points are
both clearly ``too bright'' at early epochs (see Fig. \ref{bvri}).  Similar
problems with data obtained for other SNe~Ia with the YALO 1.0~m $R$ filter
have been documented \citep[see][] {Sun00, Str_etal02};  indeed, this filter
was replaced in 2001 by a filter which provides a much better match to the
Kron-Cousins system. Problems with SN~Ia photometry taken with the YALO $B$
and $V$ filters have also been observed by \citet{Kri_etal03} and
\citet{Can_etal03}. In principle, with precise knowledge of the effective
bandpasses of the different telescope/filter/CCD combinations and optical
spectrophotometry with sufficient temporal coverage, it should be possible to
correct all of the photometry, including the YALO observations, to the
standard $BVRI$ system as defined by \citet{Bes90}.  To date, however,
attempts to do this have been only partially successful \citep{Str_etal02,
Kri_etal03, Kri_etal04c}.  Since the YALO data are not needed to plug any
critical gaps in the light curves, we can limit ourselves to a consideration
of the optical photometry based on the CTIO 0.9~m and 1.5~m telescopes.

The maximum light magnitudes in $BVRI$ and the Julian Dates that these were
attained are listed in Table \ref{maxima}.  Direct measurement of the $B$-band
light curve of SN~1999ac yields a value of 1.32 $\pm$ 0.08 for the decline
rate parameter \dmm, which is defined as the amount in magnitudes that the $B$
light curve declines during the first 15 days following maximum \citep{Phi93}.  
Correcting for a total (Galactic + host galaxy) reddening of E($B-V$) = 0.14
(see \S4.1), this value becomes \dmm\ = 1.33 
\citep[see][] {Phi_etal99} -- but as we shall see, this value 
of the decline rate does not
tell the whole truth about the light curves. As shown in Fig.~4, the shapes of
the $BVRI$ light curves are complicated, and not really a good match to the
light curves of other well-observed SNe~Ia.  Here the various light curve
templates have all been adjusted to coincide at maximum light.  Note, in
particular, the poor correspondence with the light curves of SN~1994D, whose
$\Delta$m$_{15}(B)$ value (1.32) matches that of SN~1999ac.  Only over the
first 15 days after maximum do the $B$-band light curves of these two SNe
evolve in a similar fashion; at both earlier and later epochs the light curve
of SN~1994D falls below that of SN~1999ac as if SN~1999ac were a slower
riser and slower decliner than SN~1994D.  The same is true in the $V$, $R$, and $I$ 
bands.

In spite of these indications that SN~1999ac is a slower-declining event than
the $\Delta$m$_{15}(B)$ measurement suggests, the comparisons with the slower
declining SNe~Ia in Fig. \ref{bvri_comp} are not without problems as well.  
Before maximum, SN~1999ac rose only slightly more quickly than SN~1991T.
The $V$-band light curve of SN~1992al ($\Delta$m$_{15}(B)$ = 1.11) is a
reasonable match to the observations of 1999ac.  However, data of SN~1992al in
other passbands (specifically $B$, $R$, and $I$) poorly match SN~1999ac in
the same passbands.

Note that in $I$, the secondary maximum occurs significantly earlier in SN~1994D
than in SN~1999ac, which again indicates that SN~1999ac was effectively a slower
decliner than SN~1994D \citep{Ham_etal96b}.  However, the {\em strength} of the
secondary maximum is more consistent with a faster-declining SN~Ia. Following the
recipe given by \citet{Kri_etal01}, we calculate a flux (with respect to maximum) of
the secondary maximum between 20-40 days after $B$ maximum of $\langle$ I $\rangle
_{20-40}$ = 0.475, which is good agreement with the expected value for a SN with
$\Delta$m$_{15}$(B) = 1.33.

The optical color evolution of SN~1999ac was similarly unique, as shown in Fig.
\ref{colors_comp}.  Here the data for SN~1999ac have been corrected for a Galactic
reddening of E($B-V$) = 0.046 \citep{Sch_etal98}.  The template curves with which
the observations are compared have been corrected for both Galactic and host
components of reddening as per Table~2 of \citet{Phi_etal99}. Included also for
comparison is the photometry of SN~1999aa from \citet{Kri_etal00} corrected for
Galactic reddening only.  Prior to the epoch of $B$ maximum, the $B-V$ color of
SN~1999ac did not differ significantly from those of the other SNe~Ia plotted.  
However, $\sim$5 days after maximum, SN~1999ac began to rapidly grow redder than
all of the other SNe, reaching a maximum difference in $B-V$ of several tenths of
a magnitude by 10-15~days past maximum.  By $\sim$30 days after maximum, however,
the $B-V$ color of SN~1999ac once again resembled that of several of the
comparison SNe, and continued to evolve in a similar fashion until around day 50,
when the $B-V$ color of SN~1999ac once again began to slowly grow steadily redder
than that of the other SNe.  The peculiarities in the evolution of the $V-R$
photometry of SN~1999ac are nominally similar to those observed for $B-V$, with
1999ac starting off reasonably similar to the other events, growing significantly
redder between 5-25 days after maximum, and then evolving back to a very similar
color evolution by day 30.  Except for the very first measurements obtained, the
$V-I$ evolution of SN~1999ac was redder than that of the reference SNe during
nearly the entire period covered by our observations.  We call special attention
to the fact that the optical color evolution of SN~1999ac was not at all like that
of SN~1999aa.  In spite of its spectroscopic peculiarities, the evolution of the
latter SN was much more consistent with that of the template curves shown,
particularly those corresponding to the slower decline rates.

\subsection{Near-Infrared Photometry}

The $J_s$, $H$, and $K_s$ light curves of SN~1999ac are displayed in Fig. \ref{jhk}.
As in the case of the optical light curves, the data obtained with the
various telescopes have been plotted with different symbols.  These
observations provide excellent coverage of the rise to maximum light
in the near-infrared.  In the $J_s$ band, 
this rise is followed by a nearly symmetric decline which levels off at 
$\sim$15 days after maximum was reached.  Unfortunately, a 14-day gap in 
the observations then occurs, after which the light curve is observed to 
decline again.  The impression is that, much like the $I$-band light
curve, the secondary maximum in $J_s$ was actually more of a ``plateau''
rather than a clearly-defined peak.

The maximum light magnitudes in $J_sHK_s$ and the corresponding Julian Dates are
listed in Table \ref{maxima}.  As with other well-observed SNe~Ia, the $J_sHK_s$
maxima all occurred $\sim$2-3 days before $B$-band maximum. In $H$ and $K_s$,
the initial rise to maximum is similar to that observed in $J_s$.  This is
followed in both bands by only a slight decrease in magnitude, to a plateau
$\sim$10-15 days after maximum.  Then there is a 20-day gap when only a single
$H$-band observation was obtained. By 35-40 days past the initial maximum,
both the $H$- and $K_s$-band light curves began a decline phase which was
probably initially somewhat more rapid than at later times (see particularly
the $K_s$-band observations).

Fig. \ref{jhk} also includes near-infrared photometric loci for three
spectroscopically normal objects: SN~1998bu \citep{Jha_etal99,Her_etal00},
SN~1999ee \citep{Kri_etal04b}, and SN~2001el \citep{Kri_etal03}.  The decline
rates of these three SNe are \dmm\ = 1.01 $\pm$ 0.05, 0.94 $\pm$ 0.06, and
1.13 $\pm$ 0.04, respectively. SN~1999ee was a slow decliner, but did {\em not}
exhibit the doubly ionized lines characteristic of SN~1991T-like objects
\citep {Ham_etal02}.  SNe 1998bu and 2001el were normal
mid-range decliners.  Fig. \ref{jhk} shows that the near-IR
light curves of SNe 1999ac and 2001el match reasonably well in the first 20
days after the IR maxima.  SN~1999ac was fainter at 10 days before and 40 days 
after T($B_{max}$), consistent with its larger value of \dmm.

While there are still very few SN~Ia that have well sampled near-IR light
curves, certain trends are suggested by the presently available data. Objects
with small values of \dmm\ are also slower risers/decliners in the IR. Objects
with larger values of \dmm\ are faster risers/decliners in the IR.  In the
$I$-band the slow decliners have later secondary peaks \citep{Ham_etal96b,
Rie_etal96, Nob_etal05}.  This trend seems to hold for the near-IR as well.

In Fig. \ref{bvijhk} we show the $BVIJ_sHK_s$ data of SN~1999ac within
$\pm$20 days of $B$ maximum along with
various single-band templates.  The time-since-maximum-light has been
scaled according to stretch factors derived from \dmm \citep[][Fig. 3.8] {Jha02}.  
Time dilation is also taken into account.  If \dmm\
is a representative measure of the light curve characteristics, the stretched
data should lie on top of the templates.  This is clearly not the case.
At early times the $BVI$ data are several tenths of a magnitude ``too bright''.
The $J_s$ data conform to the template at early times, but the $I$ and $J_s$ data are 
too bright after T($B_{max}$).  Only in the $K_s$ band do the data conform closely
to the template before and after maximum light.  

The $V-J_s$, $V-H$, and $V-K_s$ color evolution of SN~1999ac is shown in Fig.
\ref{vjhk_colors}.  We have also added the V {\em minus} near-IR color
loci from \citet{Kri_etal00} offset by various amounts (see \S4.1).
Because the early time $V$-band data are brighter than we would expect on the basis of
\dmm\ alone, the $V$ {\em minus} near-IR colors are much too blue at early times.  
Since the $V$ and $K_s$ data conform to the templates from 0 to 20 d after
T($B_{max}$), the most reliable color index for the purposes of determining
a color excess and extinction may well be $V-K_s$ at this epoch.

\section{Optical Spectroscopy}

Four optical spectra of SN~1999ac covering phases from 13 days before
to 41 days after $B$ maximum were obtained with the LCO 2.5~m du~Pont 
telescope and WFCCD spectrograph.  The specific dates and phases, as well
as the wavelength coverage and resolution of these spectra, are given in 
Table \ref{opt_spectra}.  The spectra are plotted in Fig. \ref{spectra}.
Note that the last spectrum of 23 April (UT) was acquired under poor 
transparency conditions and consequently suffers from low signal-to-noise.
The version of this spectrum shown in Fig. \ref{spectra} has therefore been 
smoothed to an effective resolution of ~90~\AA.

The strongest features in the spectrum of SN~1999ac obtained at $-$13 days are 
identified with \ion{Ca}{2}, \ion{Fe}{3}, and \ion{Si}{2}.  This spectrum is
compared in Fig. \ref{comp_m13d} with comparable-phase spectra of SN~1991T,
SN~1999aa, and SN~1990N.  The overall impression is that these four SNe
form an orderly sequence, with the peculiar SN~1991T at the one extreme and 
the ``normal'' SN~1990N at the other.  Note, in particular, how the 
strengths of the \ion{Ca}{2}~H\&K and \ion{Si}{2}~$\lambda$6355 lines vary
continuously from SN~1991T, where they are completely absent, to
SN~1990N where they are the strongest features in the spectrum.  This 
situation lends credence to the idea that SNe~Ia like SN~1999aa and SN~1999ac 
are transition events \citep{Li_etal01a, Branch01, Gar_etal04}.  Spectra
obtained around T($B_{max}$) of these same four SNe strengthen this impression,
as illustrated in Fig. \ref{comp_m2d}.  Once again, the strengths of the
\ion{Ca}{2}~H\&K and \ion{Si}{2}~$\lambda$6355 form a smooth sequence, with
SN~1999ac lying intermediate between SN~1999aa and SN~1990N.  By this phase,
the \ion{S}{2}~$\lambda\lambda$5454,5460 lines have appeared in both
SN~1999aa and SN~1999ac, making the peculiar nature of these events far
less obvious than at earlier epochs.  Indeed, \cite{Li_etal01a} have 
emphasized the difficulty of identifying SN~1999aa-like events in the 
absence of spectral observations obtained well before maximum light.
Fig. \ref{comp_8d} shows that by a week after $B$-band maximum, it is virtually
impossible to discriminate spectroscopically between SN~1999ac and the 
``normal'' SN~1990N.

\citet{Gar_etal05} have recently presented a detailed analysis of the 
optical spectral evolution of SN~1999ac.  Their data cover essentially the
same time period ($-$15 to +42~days with respect to $B$ maximum) as ours, 
but with more frequent temporal sampling.  Their main conclusions are similar
to those presented in the previous paragraph -- i.e., that prior to 
maximum light, the spectra of SN~1999ac resembled those of SN~1999aa, but
with stronger \ion{Ca}{2} and \ion{Si}{2} absorption, whereas after maximum 
the spectra appeared essentially normal.  The \ion{Ca}{2} H\&K lines were
found to exhibit larger than average expansion velocities from 
maximum light onward, whereas the iron lines appeared to be characterized by 
somewhat lower than average velocities.  The \ion{Si}{2}~$\lambda$6355 
expansion velocities decreased monotonically from $-$14 to +32 days, and
are amongst the slowest ever observed.  Expansion velocities measured from
the minima of the \ion{Ca}{2}~H\&K and \ion{Si}{2}~$\lambda$6355 lines in
our spectra are given in the last two columns of Table \ref{opt_spectra}.
These are in good agreement with the measurements of \citet{Gar_etal05}.
Our expansion velocity for \ion{Ca}{2}~H\&K obtained at $-$13~days suggests
that at epochs before maximum the calcium velocities were {\em not}
unusually high, although very few SNe~Ia have been observed at such early
epochs.

\citet{Gar_etal05} measured values for SN~1999ac of the
$\mathcal{R}$(\ion{Si}{2}), $v_{10}$(\ion{Si}{2}), and $\dot{v}$ parameters
\footnote[20]{$\mathcal{R}$(\ion{Si}{2}) is the ratio of the depth of
\ion{Si}{2} 5800 \AA\ to \ion{Si}{2} 6100 \AA\ absorption. It is theorized to
be driven by temperature, hence the $^{56}$Ni mass synthesized in the
explosion.  $v_{10}$(\ion{Si}{2}) is the blueshift measured in the \ion{Si}{2}
$\lambda$6355 line ten days after maximum light. $\dot{v}$ is an estimate of
the expansion velocity time derivative computed after maximum light.} that, in
combination with \dmm, \citet{Ben_etal05} recently used to identify three
subgroups of SNe~Ia.  In all three of the diagrams considered by
Benetti et al. --- \dmm\ vs.  $\mathcal{R}$(\ion{Si}{2}), \dmm\ vs.  
$v_{10}$(\ion{Si}{2}), and \dmm\ vs.  $\dot{v}$ --- SN~1999ac appears to be
either an outlier or an extreme example of ``normal'' SNe~Ia.  
\citet{Gar_etal05} pointed out that the low value of
$\mathcal{R}$(\ion{Si}{2}) = 0.098$\pm$0.030 measured for SN~1999ac is
consistent with high temperature (as manifested by the presence of strong
\ion{Fe}{3} absorption lines at early epochs), but is inconsistent with the
value of \dmm = 1.33, suggesting that for this object \dmm\ might not be a
good indicator of luminosity.  We will return to this point in \S4.2.

\section{Discussion} 

\subsection{Host Galaxy Reddening}

The location of SN~1999ac in the outer regions of its host galaxy and the lack
of strong \ion{Na}{1} D lines in the spectra (which would imply the existence of
dust along the line of sight) lead us to believe that the host galaxy reddening
of this SN should be small.  
Estimating the host galaxy reddening of SN~1999ac is not so straight forward due
to the peculiarities in the evolution of the various observed colors.  If we
follow the procedures detailed in \citet{Phi_etal99} using the measured value of
the decline rate \dmm\ = 1.32 $\pm$ 0.08, the $BVI$ data imply a host galaxy
reddening of E(B$-$V) = 0.12 $\pm$ 0.03 mag.  If the dust along the line of
sight can be characterized as having a ratio of total to selective absorption
R$_V$ $\equiv$ A$_V$/E(B$-$V) = 3.1 \citep{Sne_etal78}, then A$_V$(host) = 0.37
$\pm$ 0.09 mag and A$_V$(tot) = 0.51 $\pm$ 0.09 mag. However, it should be noted
that one of the pillars of the Phillips et al. method is the assumption that all
SNe~Ia follow essentially the same $B-V$ evolution at late times (30-90 days
after $V$-band maximum).  Fig. \ref{colors_comp} shows that this is not the case
for SN~1999ac in that the data from 30 to 90 days after $V$ maximum are best
fitted by a slope of $-0.0073 \pm 0.0004$ mag d$^{-1}$, which is clearly
shallower than the value of $-$0.0118 mag d$^{-1}$ given by
\citet{Lir95}.\footnote[21]{Two other SN~Ia whose color evolution did not allow
the unambiguous determination of color excesses were the highly peculiar
SN~2000cx \citep{Li_etal01b, Can_etal03} and SN~2002bo \citep{Ben_etal04,
Kri_etal04c}.}

If we maximize the wavelength difference of two ``well behaved'' photometric
bands, it is possible to get a better handle on the extinction towards
a star or supernova.  \citet{Kri_etal00} found that SN~Ia
with mid-range decline rates exhibited uniform V {\em minus} near-IR
colors from $-9 \lesssim t - T(B_{max}) \lesssim +27$ d, allowing a determination
of the V {\em minus} near-IR color excesses and a more robust estimate
of A$_V$.  \citet{Kri_etal04b} showed that slowly declining SN~Ia have
V {\em minus} near-IR colors roughly 0.24 mag bluer than those of mid-range
decliners.

Fig. \ref{vjhk_colors} shows that no simple vertical shifting
of our V {\em minus} near-IR templates matches the SN~1999ac data from
$-9 \lesssim t - T(B_{max}) \lesssim +27$ d.  If, however, we restrict the
fits to the first 20 days after T($B_{max}$), we obtain color excesses
of E($V-J_s$) = 0.44 $\pm$ 0.08, 
   E($V-H$) = 0.47 $\pm$ 0.06, and
   E($V-K_s$) = 0.37 $\pm$ 0.06.  
As discussed above in \S2.3, the best behaved photometric band before and
after the time of maximum light was $K_s$.  Since the $V$-band data in the
first weeks after maximum is also well behaved, we would suggest that the most
reliable color index from which to derive a color excess and extinction
is $V-K_s$.  With A$_V$ = 1.129 $\pm$ 0.029 $\times$ E($V-K_s$) based on
the reddening model of \citet{Car_etal89},
%\footnote[22]{Equation 4 of \citet{Kri_etal04b}.} 
we obtain A$_V$(tot) = 0.42 $\pm$ 0.07 mag.  This implies
E($B-V$)$_{host}$ = 0.09, slightly less than we obtained from optical photometry
only.  We shall adopt this smaller value as the best estimate of the host
galaxy color excess.

According to \citet{The_etal98} the heliocentric radial velocity of the host of
SN~1999ac is 2847 km s$^{-1}$.  Corrected to the frame of the Cosmic Microwave
Background radiation, this velocity is 2942 km s$^{-1}$.  With a Hubble constant of 72
km s$^{-1}$ Mpc$^{-1}$ \citep{Fre_etal01} we get a distance modulus of $m-M$ = 33.06
$\pm$ 0.22 mag, where the uncertainty corresponds to a random velocity of $\pm$300 km
s$^{-1}$.  If $V_{max}$ = 14.20 $\pm$ 0.02 and A$_V$ = 0.42 $\pm$ 0.07 mag, the
absolute magnitude of SN~1999ac is $-$19.28 $\pm$ 0.24.  This is 0.24 mag brighter
than the value given by \citet{Li_etal03}, who assumed zero host galaxy reddening on
the basis of the position of the SN in its host and the lack of Na I D absorption in
the spectra (Li 2005, private communication).  The absolute magnitude at maximum can
be compared to $M_V = -19.13$ for a typical SN~Ia with \dmm\ = 1.33 and H$_0$ = 72
\citep[][Eqn. 18]{Phi_etal99}.

The absolute magnitudes of SN~1999ac depend critically on the adopted 
host galaxy extinction.  While this problem is potentially unsolveable
in the case of optical absolute magnitudes, serious
systematic errors in the extinction corrections are less problematic
in the near-IR.  Table 1 and Fig. 3 of \citet{Kri_etal04a} adopt total
$VJHK$ extinctions of 0.51, 0.14, 0.10, and 0.06 mag, respectively, and show
that the near-IR absolute magnitudes of SN~1999ac are statistically
equal to the mean of SN~Ia with slower decline rates.

\subsection{Luminosity and Bolometric Light Curve}

In Fig. \ref{bol} we show the bolometric light curves of SNe~1994D, 1999aa, and
1999ac, calculated using the $UBVRI$ photometry of \citet{Jha02}, and our $BVRI$
photometry.  For details on the method see \cite{Sun03}. As stated above, we assumed
a distance modulus $m-M$ = 33.06 mag and E($B-V$)$_{host}$ = 0.09 mag for SN~1999ac.  
The three objects represented in Fig. \ref{bol} have essentially the same peak
bolometric luminosity.  While SNe~1994D and 1999ac have the same observed $B$-band
decline rates, their bolometric decline rates are clearly different.

\citet{Can_etal03} considered the difference of log L at maximum compared to log L at
90 days after maximum.  Calling this parameter $\Delta$log$_{10}$ L$_{90}$,
\citet{Can_etal03} found that it correlates with \dmm, which allowed
them to gain insight
into the nature of the effective decline rate of the peculiar SN~2000cx.  Both
SNe~1999aa and 1999ac have $\Delta$log$_{10}$ L$_{90}$ $\approx$ 1.40.
Taken altogether, the evidence suggests that SN~1999ac was energetically
much more similar to the
spectroscopically peculiar slow decliner SN~1999aa than to the faster
declining SN~1994D.  

Finally, in Fig. \ref{bol_drr} we show the equivalent of the luminosity
vs. decline rate relation
for SN~Ia, but using bolometric light curves.  The ordinate is the
logarithm of the peak luminosity, while the abscissa represents the 
decline in log L over the first 15 days after the bolometric maximum.
SN~1999ac had a peak bolometric luminosity and bolometric decline rate
near the mid-range of most of the objects represented.  Except for
the record-setting slow decliner SN~2001ay \citep{Nug_etal06} and
the faster declining SNe~1992A \citep{Ham_etal96b} and 
1999by \citep{Garn_etal04}, SNe~Ia appear to occupy a
rather small parameter space in the bolometric decline rate graph.
This argues in favor of a uniform explosion mechanism for the majority
of Type Ia supernovae.

% I CAN'T THINK OF ANYTHING SIGNIFICANT TO SAY ABOUT 99ac FROM THE NICKEL-56
% YIELDS CALCULATED BY STRITZINGER ET AL.  A NUMBER OF OBJECTS HAVE QUITE
% SIMILAR NICKEL YIELDS.  
%
%On the basis of their own derived bolometric light curve analysis \citet{Str_etal05}
%derived a $^{56}$Ni mass of 0.67 M$_{\sun}$ for SN~1999ac, similar to that of
%SNe~1989B, 1994D, 1995D, and 1998aq, but considerably less than that of the
%prototypical slow decliner SN~1991T (0.93 M$_{\sun}$).

\section{Conclusions} 

The Type Ia supernova 1999ac was peculiar in several respects.  As shown by
\citet{Gar_etal05} and by our data, the spectra of SN~1999ac resembled the
unusual SN~1999aa at early times.  From maximum light onward, however, the 
spectrum was essentially normal.  This underscores a thorny issue of SN
classification: if early-time spectra are not obtained, there may be no way to
determine if a particular Type Ia supernova is unusual 
\citep[see also][] {Li_etal01a}.

The optical photometry of SN~1999ac was unusual in that it had
a slow rise rate (like SN~1991T), but a much more rapid $B$-band decline rate
(like SN~1994D). In this sense it was the opposite of SN~2000cx, which was a
rapid riser, but slow decliner.  This is in striking contrast to the 
$B$- and $V$-band light curves of most SN~Ia which can be 
characterized by a single stretch factor \citep{Gol_etal01}.
In the case of SN~1999ac, our best estimate of the extinction is obtained
from the $V-K_s$ colors during the first 20 days 
after T($B_{max}$).  We find A$_V$(tot) = 0.42 mag, implying a
non-zero host galaxy reddening of E($B-V$)$_{host}$ = 0.09 mag.

In spite of the peculiar photometric properties of SN~1999ac, because near-infrared
extinction corrections are an order of magnitude smaller than at optical wavelengths,
the infrared absolute magnitudes of SN~1999ac cannot be systematically in error by
more than a few hundredths of a magnitude. The derived infrared absolute magnitudes
\citep{Kri_etal04a} are statistically equal to the mean of all other SN~Ia with slower
decline rates.  As a result, the unusual SN~1999ac can still be regarded as an
infrared standard candle.  Its $J_sHK_s$ light curves were very similar to those of 
the normal SN~2001el.

The bolometric light curve of SN~1999ac closely resembled that of the
spectroscopically peculiar slow decliner SN~1999aa.  The differences of their
bolometric luminosities at maximum compared to 90 days after maximum are almost
identical.  As measured by this parameter, SN~1999ac behaved as if it were 
a slow decliner.  Hence, in terms of its energetics and pre-maximum 
spectral characteristics, there is little doubt that SN~1999ac was closely
related to other peculiar, slow-declining events such as SNe~1999aa and
1991T.

Finally, we have also examined the luminosity vs. decline rate relation for 
SN~Ia, but considered
according to their bolometric light curves.  Most events occupy a rather
small parameter space, suggesting a uniform
explosion mechanism for these supernovae.

\vspace {1 cm}

\acknowledgments

M. Hamuy and G. Galaz acknowledge the support of Fondap Center for
Astrophysics grant number 15010003. 
Some of the observations were obtained with the Yale 1 m telescope at CTIO,
operated by the Yale/AURA/Lisbon/OSU (YALO) consortium, now the Small
and Moderate Aperture Research Telescope System (SMARTS) Consortium.
We acknowledge funding from STScI from the grants GO-8243.02-A and
GO-8648.10-A.  We made use of data in the NASA/IPAC
Extragalactic Database (NED).  We thank Max Stritzinger for discussions
relating to bolometric light curves, and Weidong Li for facilitating his
published spectra of SN~1999aa.

\begin{deluxetable}{ccccccccc}
\tabletypesize{\scriptsize}
\tablewidth{0pc}
\tablecaption{Optical Photometric Sequence near SN 1999ac\tablenotemark{a}\label{standards}}
\tablehead{   \colhead{Star ID\tablenotemark{b}} &
\colhead {V} & \colhead{$B-V$} &
\colhead{$V-R_{KC}$} & \colhead{$V-I_{KC}$} &
\colhead{N$_{V}$} & \colhead{N$_{B-V}$} & \colhead{N$_{V-R_{KC}}$} & 
\colhead{N$_{V-I_{KC}}$} } 
\startdata
1  & 15.839 (0.004) & 1.119 (0.009) & 0.615 (0.005) & 1.185 (0.005) & 11 & 12	& 12	& 12 \\
2  & 17.359 (0.007) & 1.485 (0.035) & 1.001 (0.008) & 2.102 (0.009) & 9	 & 10	& 10	& 10 \\
3  & 16.193 (0.004) & 0.747 (0.010) & 0.412 (0.007) & 0.789 (0.007) & 11 & 12	& 12	& 12 \\
4  & 16.889 (0.007) & 0.662 (0.012) & 0.415 (0.010) & 0.801 (0.009) & 10 & 11	& 11	& 12 \\
5  & 13.776 (0.002) & 0.739 (0.003) & 0.403 (0.003) & 0.759 (0.003) & 11 & 12	& 12	& 12 \\
6  & 17.814 (0.008) & 0.992 (0.027) & 0.571 (0.013) & 1.104 (0.011) & 7	 & 9	& 10	& 10 \\
7  & 14.767 (0.002) & 0.680 (0.006) & 0.388 (0.004) & 0.757 (0.004) & 10 & 11	& 11	& 11 \\
8  & 17.222 (0.007) & 0.672 (0.016) & 0.426 (0.009) & 0.839 (0.011) & 8	 & 10	& 10	& 10 \\
9  & 17.414 (0.009) & 0.627 (0.014) & 0.362 (0.015) & 0.745 (0.011) & 9	 & 10	& 10	& 10 \\
%
% Object #10 IS a star, not a galaxy.  Why is it not in Pablo's list of field stars?
%
11 & 18.056 (0.016) & 0.709 (0.039) & 0.426 (0.023) & 0.821 (0.022) & 4	 & 8	& 8	& 8 \\
12 & 18.730 (0.020) & 0.951 (0.077) & 0.552 (0.029) & 1.117 (0.032) & 2	 & 4	& 4	& 4 \\
13 & 17.920 (0.021) & 1.451 (0.034) & 0.892 (0.025) & 1.821 (0.025) & 2	 & 3	& 3	& 3 \\
%
% There are a few off the bottom of Lou's new chart (#14, 15, 16, 18).
%
%14 & 16.312 (0.009) & 0.977 (0.018) & 0.544 (0.018) & 1.022 (0.017) & 2	 & 2	& 2	& 2 \\
%15 & 15.727 (0.006) & 0.653 (0.019) & 0.387 (0.009) & 0.763 (0.007) & 3	 & 3	& 3	& 3 \\
%16 & 18.422 (0.035) & 1.266 (0.048) & 0.811 (0.037) & 1.493 (0.040) & 3	 & 3	& 3	& 3 \\
17 & 17.896 (0.016) & 0.625 (0.025) & 0.369 (0.037) & 0.741 (0.022) & 2	 & 2	& 2	& 2 \\
%18 & 16.302 (0.013) & 0.776 (0.049) & 0.428 (0.019) & 0.841 (0.014) & 2	 & 2	& 2	& 2 \\
19 & 17.259 (0.020) & 0.794 (0.027) & 0.455 (0.022) & 0.906 (0.025) & 3	 & 3	& 3	& 3 \\
20 & 17.990 (0.010) & 0.771 (0.018) & 0.444 (0.016) & 0.890 (0.014) & 7	 & 7	& 7	& 7 \\
21 & 18.565 (0.025) & 1.629 (0.092) & 1.168 (0.029) & 2.671 (0.027) & 2	 & 4	& 4	& 4 \\
22 & 17.165 (0.006) & 0.830 (0.024) & 0.452 (0.009) & 0.909 (0.010) & 6	 & 6	& 6	& 6 \\
23 & 17.925 (0.018) & 0.559 (0.027) & 0.344 (0.022) & 0.706 (0.023) & 4	 & 4	& 4	& 4 \\
24 & 17.989 (0.016) & 0.798 (0.104) & 0.504 (0.035) & 0.984 (0.031) & 2	 & 2	& 2	& 2 \\
\enddata
\tablenotetext{a} {In this and subsequent tables the numbers in parentheses are
the 1$\sigma$ uncertainties.}
\tablenotetext{b} {The identifications are the same as in Fig. \ref{finder}.}
\end{deluxetable}

\begin{deluxetable}{ccccccl}
\tabletypesize{\scriptsize}
\tablewidth{0pc}
\tablecaption{$BV(RI)_{KC}$ Photometry of SN 1999ac\label{opt_photom}}
\tablehead{   \colhead{JD$-$2,450,000} &
\colhead{UT Date$^a$} & \colhead {$B$} & \colhead{$V$} &
\colhead{$R_{KC}$} & \colhead{$I_{KC}$} &
\colhead{Observer+Telescope} }
\startdata
 1238.90 &  Mar01.40 & 15.485 (0.013) & 15.487 (0.008) & 15.444 (0.013) & 15.415 (0.016) & Germany CTIO 0.9~m   \\
 1239.90 &  Mar02.40 & 15.285 (0.012) & 15.294 (0.007) & 15.255 (0.012) & 15.228 (0.015) & Germany CTIO 0.9~m  \\
 1240.89 &  Mar03.40 & 15.106 (0.015) & 15.125 (0.009) & 15.082 (0.020) & 15.055 (0.023) & Germany CTIO 0.9~m  \\
 1241.87 &  Mar04.37 & 15.039 (0.026) & 14.901 (0.014) &     \nodata    & 14.802 (0.027) & Espinoza/Gonzalez CTIO YALO 1~m \\   
 1243.88 &  Mar06.38 & 14.763 (0.028) & 14.636 (0.009) &     \nodata    & 14.521 (0.021) & Espinoza/Gonzalez CTIO YALO 1~m    \\

 1246.85 &  Mar09.35 & 14.480 (0.033) & 14.336 (0.022) &     \nodata    & 14.233 (0.053) & Espinoza/Gonzalez CTIO YALO 1~m     \\
 1252.85 &  Mar15.35 & 14.386 (0.020) & 14.136 (0.012) &     \nodata    & 14.286 (0.018) & Espinoza/Gonzalez CTIO YALO 1~m     \\
 1257.88 &  Mar20.38 & 14.689 (0.014) & 14.326 (0.001) & 14.248 (0.006) & 14.523 (0.004) & Strolger CTIO 0.9~m      \\
 1263.83 &  Mar26.33 & 15.418 (0.028) & 14.618 (0.015) &     \nodata    & 14.706 (0.023) & Espinoza/Gonzalez CTIO YALO 1~m \\  
 1266.88 &  Mar29.38 & 15.739 (0.028) & 14.888 (0.011) & 14.678 (0.019) & 14.783 (0.016) & Strolger CTIO 0.9~m   \\

 1267.91 &  Mar30.41 & 15.841 (0.025) & 14.952 (0.009) & 14.715 (0.016) & 14.764 (0.016) & Strolger CTIO 0.9~m   \\
 1274.89 &  Apr06.39 & 16.353 (0.031) & 15.268 (0.014) & 14.890 (0.023) & 14.792 (0.022) & Strolger CTIO 0.9~m  \\
 1275.86 &  Apr07.36 & 16.355 (0.023) & 15.297 (0.007) & 14.904 (0.011) & 14.812 (0.016) & Strolger CTIO 0.9~m \\
 1277.86 &  Apr09.36 & 16.512 (0.020) & 15.386 (0.007) & 14.987 (0.017) & 14.821 (0.017) & C.Smith/Strolger CTIO 1.5~m \\ 
 1284.78 &  Apr16.29 & 16.917 (0.025) & 15.824 (0.017) & 15.415 (0.028) & 15.104 (0.036) & M.T.Ruiz CTIO 0.9~m    \\

 1285.84 &  Apr17.34 & 17.001 (0.034) & 15.912 (0.009) &     \nodata    & 15.164 (0.015) & Krisciunas/Hastings APO 3.5m    \\
 1287.81 &  Apr19.31 & 17.088 (0.021) & 16.023 (0.013) & 15.623 (0.023) & 15.316 (0.027) & Germany CTIO 0.9~m    \\
 1291.88 &  Apr23.38 & 17.222 (0.088) & 16.194 (0.050) & 15.834 (0.097) & 15.565 (0.076) & Strolger/C.Smith/R.Smith CTIO 1.5~m \\   
 1294.76 &  Apr26.26 & 17.300 (0.055) & 16.287 (0.017) &     \nodata    & 15.689 (0.046) & Espinoza/Gonzalez CTIO YALO 1~m \\
 1296.73 &  Apr28.23 & 17.361 (0.019) & 16.362 (0.012) & 16.062 (0.022) & 15.807 (0.020) & Strolger/C.Smith/Bonati CTIO 1.5~m   \\    

 1297.71 &  Apr29.21 & 17.327 (0.030) & 16.357 (0.012) &     \nodata    & 15.840 (0.041) & Espinoza/Gonzalez CTIO YALO 1~m  \\
 1300.60 &  May02.10 & 17.460 (0.070) & 16.403 (0.029) &     \nodata    & 15.921 (0.093) & Espinoza/Gonzalez CTIO YALO 1~m \\
 1304.79 &  May06.29 & 17.523 (0.016) & 16.562 (0.009) & 16.298 (0.014) & 16.134 (0.015) & Strolger CTIO 1.5~m       \\
 1305.82 &  May07.32 & 17.511 (0.005) & 16.581 (0.002) & 16.338 (0.006) & 16.185 (0.010) & Strolger CTIO 1.5~m       \\
 1306.83 &  May08.33 & 17.528 (0.032) & 16.586 (0.017) &     \nodata    & 16.146 (0.043) & Espinoza/Gonzalez CTIO YALO 1~m \\      

 1313.83 &  May15.33 & 17.670 (0.028) & 16.788 (0.004) & 16.605 (0.008) &     \nodata    & Krisciunas/McMillan APO 3.5 \\
 1314.75 &  May16.25 & 17.683 (0.017) & 16.801 (0.008) & 16.608 (0.015) &     \nodata    & Germany CTIO 0.9~m      \\
 1315.72 &  May17.22 & 17.688 (0.023) & 16.817 (0.010) &     \nodata    & 16.473 (0.021) & Espinoza/Gonzalez CTIO YALO 1~m \\
 1316.83 &  May18.33 & 17.711 (0.020) & 16.855 (0.012) & 16.688 (0.022) & 16.554 (0.018) & Krisciunas/McMillan APO 3.5m  \\
 1318.68 &  May20.19 & 17.755 (0.028) & 16.863 (0.018) &     \nodata    & 16.609 (0.029) & Espinoza/Gonzalez CTIO YALO 1~m \\   

 1329.68 &  May31.18 & 17.892 (0.038) & 17.088 (0.017) &     \nodata    & 16.909 (0.046) & Espinoza/Gonzalez CTIO YALO 1~m    \\ 
 1336.64 &  Jun07.14 & 17.997 (0.024) & 17.266 (0.011) &     \nodata    & 17.192 (0.044) & Espinoza/Gonzalez CTIO YALO 1~m      \\
 1363.60 &  Jul04.10 & 18.460 (0.025) & 17.940 (0.013) & 18.118 (0.025) & 18.169 (0.039) & Germany CTIO 0.9~m  \\
 1364.57 &  Jul05.07 & 18.439 (0.032) & 17.952 (0.019) & 18.105 (0.034) & 18.214 (0.044) & Germany CTIO 0.9~m  \\
 1365.60 &  Jul06.10 & 18.489 (0.021) & 17.973 (0.012) & 18.130 (0.021) & 18.204 (0.029) & Germany CTIO 1.5~m     \\  

 1375.51 &  Jul16.01 & 18.618 (0.031) & 18.184 (0.018) &     \nodata    &    \nodata     & Corwin CTIO 0.9~m       \\
 1379.50 &  Jul20.00 & 18.698 (0.044) & 18.240 (0.024) & 18.469 (0.041) & 18.540 (0.052) & Corwin CTIO 0.9~m       \\
 1381.52 &  Jul22.02 & 18.722 (0.040) & 18.273 (0.022) & 18.518 (0.038) &     \nodata    & Corwin CTIO 0.9~m       \\
 1382.49 &  Jul22.99 & 18.724 (0.044) & 18.306 (0.024) & 18.565 (0.040) & 18.570 (0.053) & Corwin CTIO 0.9~m       \\
\enddata
\tablenotetext{a} {Year is 1999.}
%$^b$Note b.
%$^c$Note c.}
\end{deluxetable}

\begin{deluxetable}{cccc}
\tablewidth{0pc}
\tablecaption{Infrared Photometric Sequence near SN 1999ac\label{ir_stds}}
\tablehead{   \colhead{Star ID$^a$} &
\colhead{$J_s$} & \colhead{$H$} & \colhead{$K_s$} } 
\startdata
1    & 13.804 (0.003) & 13.183 (0.005) & 13.087 (0.008) \\
2    & 14.126 (0.008) & 13.482 (0.016) & 13.256 (0.015) \\
6    & 16.167 (0.049) & 15.694 (0.043) & 15.585 (0.086) \\
8    & 15.869 (0.013) & 14.440 (0.013) & 15.403 (0.015) \\
9    & 16.303 (0.040) & 16.061 (0.039) & 16.010 (0.029) \\
IR3  & 15.966 (0.005) & 15.427 (0.013) & 15.192 (0.017) \\
IR7  & 17.093 (0.026) & 16.540 (0.028) & 16.299 (0.069) \\
\enddata
\tablenotetext{a} {The identifications are 
the same as those in Table \ref{standards} and Fig. \ref{finder}.}
%$^b$Note b.
%$^c$Note c.}
\end{deluxetable}

\begin{deluxetable}{cccccl}
\tabletypesize{\scriptsize}
\tablewidth{0pc}
\tablecaption{Near Infrared Photometry of SN 1999ac\label{ir_photom}}
\tablehead{   \colhead{JD$-$2,450,000} & \colhead{UT Date$^a$} &
\colhead {$J_s$} & \colhead{$H$} &
\colhead{$K_s$} & \colhead{Observer+Telescope} }
\startdata
1237.89	& Feb28.39 & 15.919 (0.019) & 15.982 (0.036) & 16.040 (0.059) & Galaz LCO 1~m \\
1239.83	& Mar02.33 & 15.461 (0.040) & 15.575 (0.048) & 15.497 (0.062) & Galaz LCO 1~m \\
1242.90	& Mar05.40 & 14.942 (0.015) & 15.042 (0.019) & 14.900 (0.028) & Galaz LCO 1~m \\
1243.90	& Mar06.40 & 14.819 (0.015) & 14.877 (0.017) & 14.806 (0.024) & Galaz LCO 1~m \\
1244.90	& Mar07.40 & 14.673 (0.020) & 14.776 (0.021) & 14.773 (0.026) & Galaz LCO 1~m \\

1246.89	& Mar09.39 & 14.509 (0.021) & 14.702 (0.027) & 14.527 (0.027) & Galaz LCO 1~m \\
1247.91	& Mar10.41 & 14.514 (0.020) & 14.672 (0.031) & 14.536 (0.026) & Galaz LCO 1~m \\
1252.83	& Mar15.33 & 14.567 (0.019) & 14.795 (0.022) & 14.621 (0.024) & Phillips LCO 1~m \\
1253.84	& Mar16.34 & 14.736 (0.021) & 14.831 (0.023) & 14.638 (0.038) & Phillips LCO 1~m \\
1254.84	& Mar17.34 & 14.781 (0.018) & 14.871 (0.023) & 14.652 (0.028) & Phillips LCO 1~m \\

1255.86	& Mar18.36 & 14.845 (0.020) & 14.803 (0.036) & 14.690 (0.030) & Roth LCO 1~m \\
1256.79	& Mar19.29 & 14.946 (0.018) & 14.837 (0.023) & 14.720 (0.024) & Roth LCO 1~m \\
1257.84	& Mar20.34 & 15.061 (0.015) & 14.777 (0.027) & 14.726 (0.037) & Roth LCO 1~m \\
1258.82	& Mar21.32 & 15.210 (0.021) & 14.809 (0.031) & 14.761 (0.031) & Roth LCO 1~m \\
1259.85	& Mar22.35 & 15.329 (0.028) & 14.863 (0.029) & 14.767 (0.035) & Roth LCO 1~m \\

1260.78	& Mar23.28 & 15.456 (0.032) & 14.835 (0.034) & 14.777 (0.034) & Roth LCO 1m \\
1262.84	& Mar25.34 & 15.599 (0.043) & 14.770 (0.026) & 14.796 (0.026) & Galaz LCO 2.5~m \\
1264.88	& Mar27.38 & 15.738 (0.020) & 14.759 (0.019) & 14.807 (0.020) & Galaz LCO 2.5~m\\
1266.90	& Mar29.40 & 15.757 (0.023) & 14.742 (0.021) & 14.727 (0.039) & Galaz LCO 2.5~m\\
1267.78	& Mar30.28 & 15.742 (0.020) & 14.746 (0.028) & 14.749 (0.034) & Phillips LCO 2.5~m \\

1270.82	& Apr02.32 & 15.753 (0.022) &     \nodata    &   \nodata      & Muena LCO 1~m \\
1272.85	& Apr04.35 & 15.858 (0.008) &     \nodata    &   \nodata      & Marzke/Persson/Beckett LCO 2.5~m \\ 
1273.84	& Apr05.34 &     \nodata    & 14.845 (0.020) &   \nodata      & Marzke/Persson/Beckett LCO 2.5~m \\
1287.83	& Apr19.33 & 16.055 (0.028) & 15.377 (0.040) & 15.495 (0.042) & Galaz/Hamuy LCO 1~m \\
1288.87	& Apr20.37 & 16.085 (0.018) & 15.396 (0.025) & 15.669 (0.050) & Hamuy LCO 1~m \\

1289.86	& Apr21.36 & 16.170 (0.028) & 15.359 (0.025) & 15.502 (0.118) & Hamuy LCO 1~m \\
1290.82	& Apr22.32 & 16.254 (0.032) & 15.488 (0.065) & 15.737 (0.051) & Phillips LCO 1~m \\
1291.77	& Apr23.27 & 16.356 (0.026) & 15.561 (0.039) & 15.984 (0.065) & Roth LCO 1~m \\
1294.81	& Apr26.31 & 16.539 (0.036) & 15.724 (0.029) & 15.899 (0.040) & Roth LCO 1~m \\
1298.66	& Apr30.16 &     \nodata    & 15.894 (0.053) &   \nodata      & Ivanov SO 1.5~m \\

1300.78	& May02.28 &     \nodata    & 15.969 (0.015) &   \nodata      & McCarthy/Abraham LCO 2.5~m \\
1303.71	& May05.21 &     \nodata    &    \nodata     & 16.293 (0.040) & Ivanov SO 2.3~m \\
1313.78 & May15.28 & 18.024 (0.037) &    \nodata     &   \nodata      & Firth/Roth/McMahon LCO 2.5~m \\ 
1330.63	& Jun01.13 &     \nodata    &    \nodata     & 17.231 (0.073) & Hungerford SO 2.3~m \\
\enddata
\tablenotetext{a} {Year is 1999.}
%$^b$Note b.
%$^c$Note c.}
\end{deluxetable}

\begin{deluxetable}{ccc}
\tablewidth{0pc}
\tablecaption{Observed Maximum Light Data for SN~1999ac\label{maxima}}
\tablehead{   \colhead{Filter} &
\colhead{Epoch (J.D. 2450000+)} & \colhead{Magnitude} }
\startdata
$B$    & 1251.0 (0.5) & 14.27 (0.02) \\
$V$    & 1251.9 (0.5) & 14.20 (0.02) \\
$R$    & 1252.0 (0.5) & 14.17 (0.02) \\
$I$    & 1249.1 (0.5) & 14.29 (0.02) \\
$J_s$  & 1249.2 (0.5) & 14.51 (0.04) \\
$H$    & 1248.0 (0.5) & 14.67 (0.04) \\
$K_s$  & 1248.5 (1.0) & 14.54 (0.04) \\
\enddata
\end{deluxetable}

\begin{deluxetable}{cclcccc}
\tabletypesize{\scriptsize}
\tablewidth{0pc}
\tablecaption{Optical Spectroscopy of SN 1999ac\label{opt_spectra}}
\tablehead{ \multicolumn{3}{c}{} & \multicolumn{2}{c}{Wavelength} &
\multicolumn{2}{c}{Expansion Velocity (km s$^{-1}$)} \\ 
\colhead{JD$-$2,450,000} &
\colhead{UT Date$^a$} & \colhead {Phase$^b$} & 
\colhead{Coverage (\AA)} & \colhead{Resolution (\AA)} & 
\colhead{Ca~II$^c$} & \colhead{Si~II$^d$} }
\startdata
 1237.89 &  Feb28.39 & \phs$-$13 & 3700-9200 &  7 & 18856 (500) & 12831 (200) \\
 1248.89 &  Mar11.39 & \phs$-$2 & 3580-9270 &  7 & 14626 (100) & 10747 (100) \\
 1258.88 &  Mar21.38 & \phs+8 & 3600-9270 &  7 & 14189 (100) &  9219 (100) \\
 1291.85 &  Apr23.35 & \phs+41 & 3670-9230 &  7 & 11126 (500) &     \nodata \\
\enddata
\tablenotetext{a} {Year is 1999.}
\tablenotetext{b} {Number of days with respect to epoch of $B$-band maximum.}
\tablenotetext{c} {Rest frame wavelength = 3945.1 \AA.}
\tablenotetext{d} {Rest frame wavelength = 6355.2 \AA.}
\end{deluxetable}

\clearpage

\figcaption[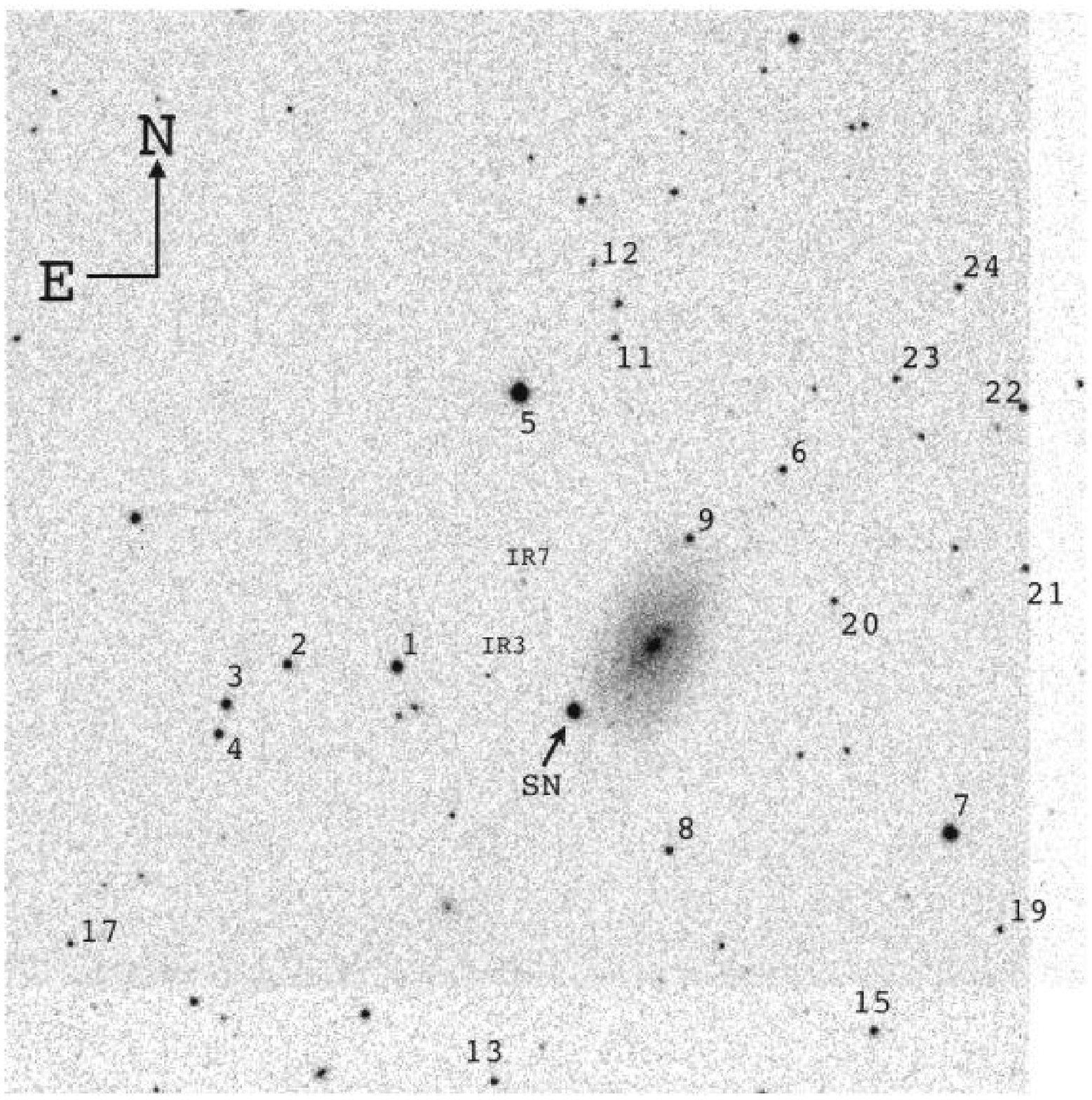]
{NGC 6063, SN~1999ac, and the field stars nearby.  This 7.3 by 7.3 arcmin
image was made from multiple $V$-band exposures obtained with the CTIO 0.9~m 
telescope shortly after the time of maximum light of the SN. 
\label{finder}
}

\figcaption[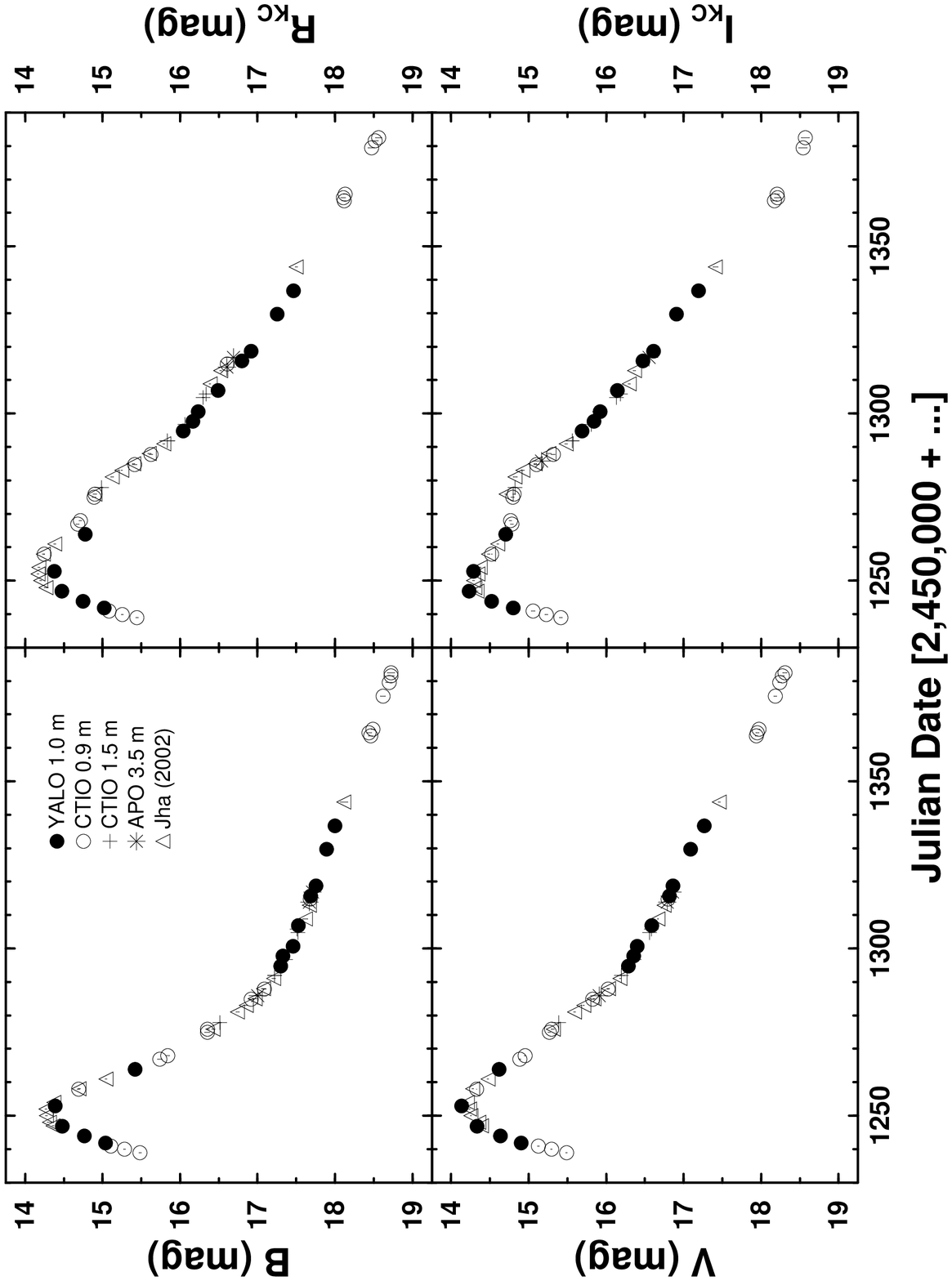]
{$BVRI$ photometry of SN 1999ac.  Data for the 
various telescopes used are plotted with different symbols, as is 
also the photometry of \citet{Jha02}.
\label{bvri}
}

\figcaption[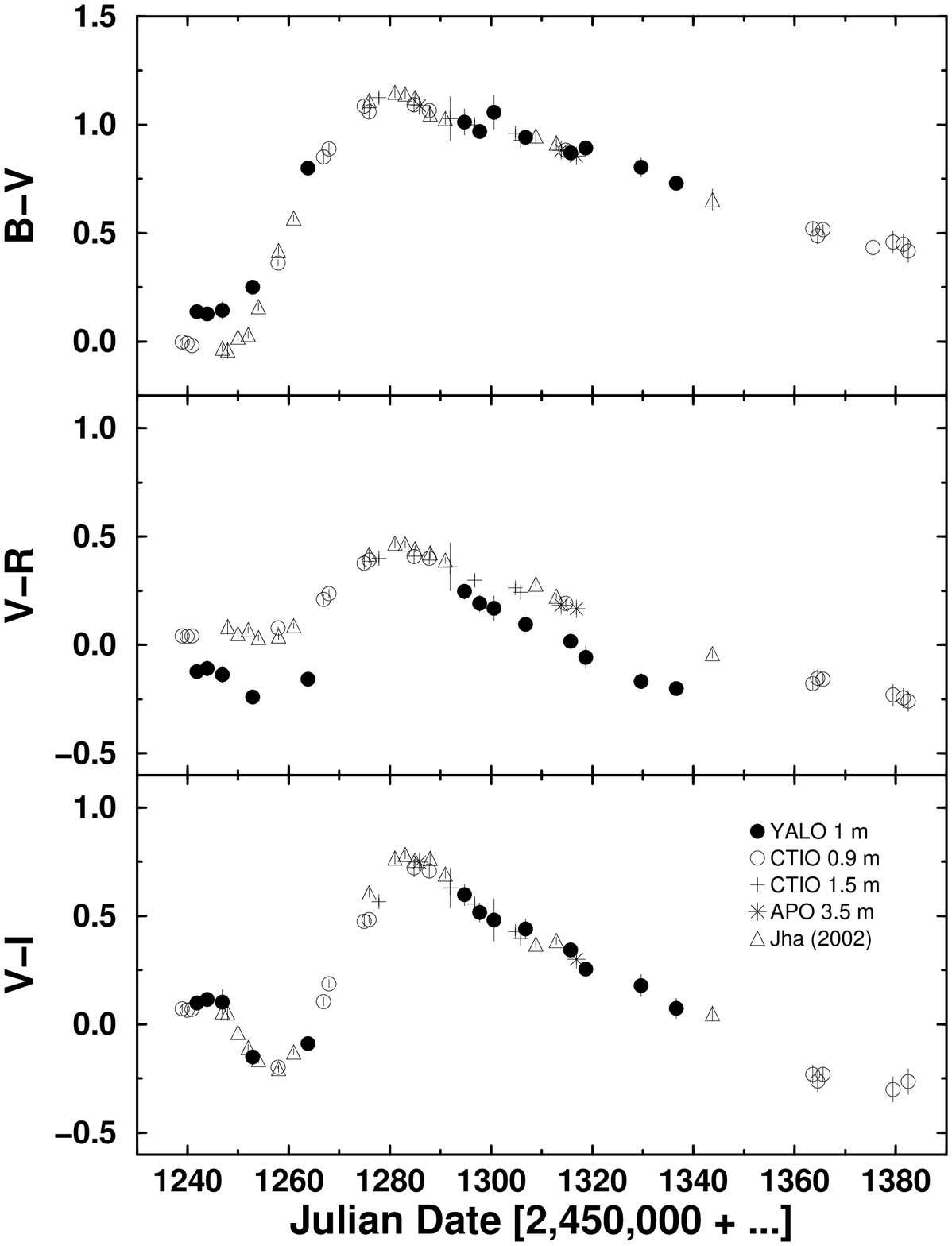]
{Top Panel -- Observed $B-V$, $V-R$, and $V-I$ colors of
SN~1999ac.  Data for the various telescopes used are plotted  
with different symbols, as is also the photometry of \citet{Jha02}.
\label{colors}
}

\figcaption[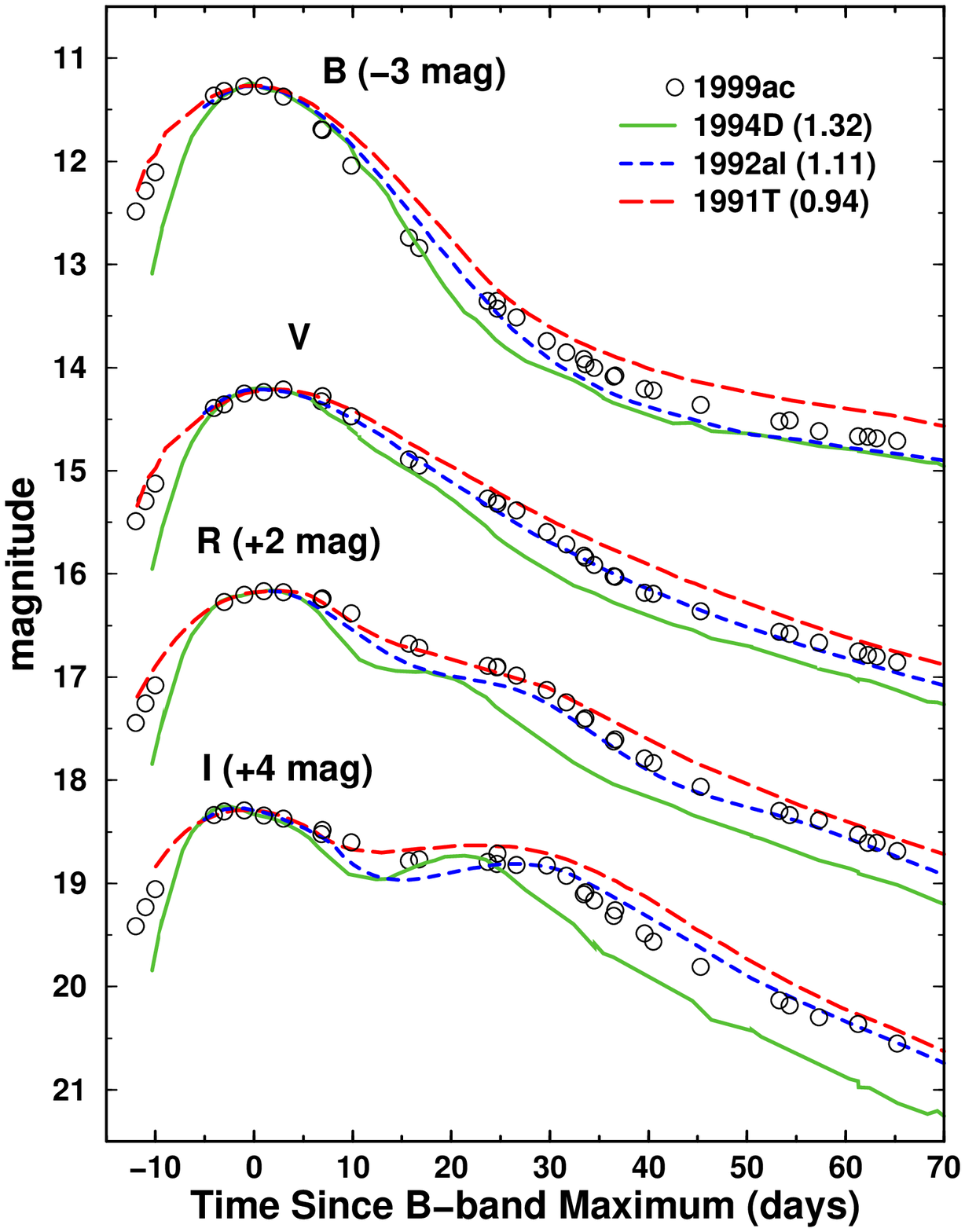]
{BVRI photometry of SN 1999ac compared with the light
curves of the Type Ia SNe 1994D ($\Delta$m$_{15}(B)$ = 1.32),
1991T ($\Delta$m$_{15}(B)$ = 0.94), and 1992al ($\Delta$m$_{15}(B)$ = 1.11).
The curves have been adjusted to be the same at maximum
in each filter.  Note that while SN 1999ac and SN 1994D
have very similar values (1.33 vs. 1.32) of the decline
rate parameter $\Delta$m$_{15}(B)$, their light curves are not at  
all identical.  Even more striking is the very slow
rise to maximum of SN 1999ac as compared to SN 1994D.
\label{bvri_comp}
}

\figcaption[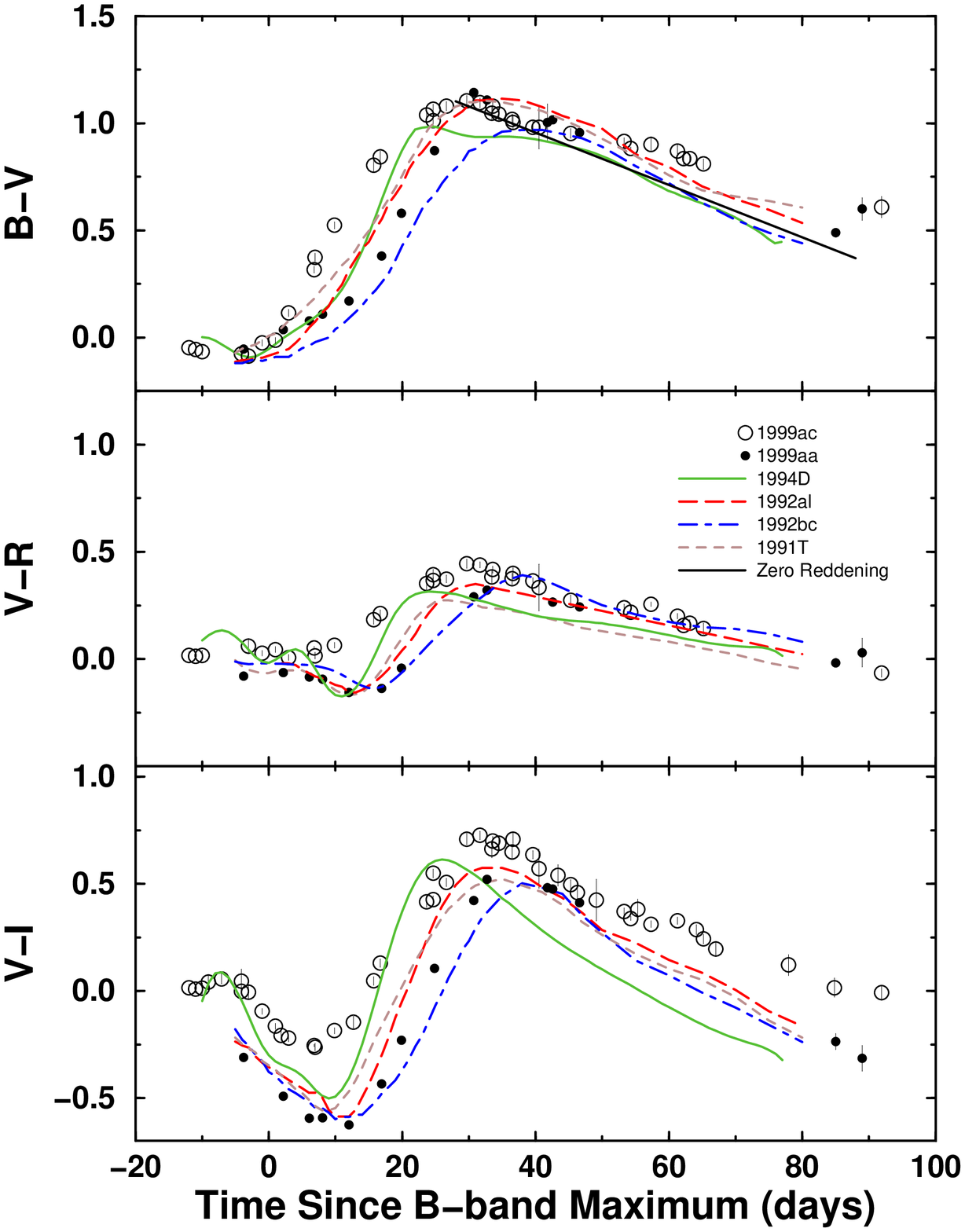]
{$B-V$, $V-R$, and $V-I$ colors of SN 1999ac corrected for a
color excess E($B-V$) = 0.046 mag due to dust in our Galaxy
(Schlegel, Finkbeiner, \& Davis 1998) and plotted as a function of the
time (in days) since the date of $B$ maximum (see Table~5).
Plotted for comparison are similar data for SN~1999aa (Krisciunas et al.
2000) corrected for a Galactic reddening of E($B-V$) = 0.040
(Schlegel, Finkbeiner, \& Davis 1998).  Shown schematically is the
evolution in these three colors of SNe~1994D ($\Delta$m$_{15}(B)$ = 1.32), 
1991T ($\Delta$m$_{15}(B)$ = 0.94), and 1992al ($\Delta$m$_{15}(B)$ = 1.11)
corrected for the Galactic and host galaxy reddenings given in Table~2
of Phillips et al. (1999).  The straight black line shows the \citet{Lir95}
relation for unreddened SNe~Ia.
\label{colors_comp}
}

\figcaption[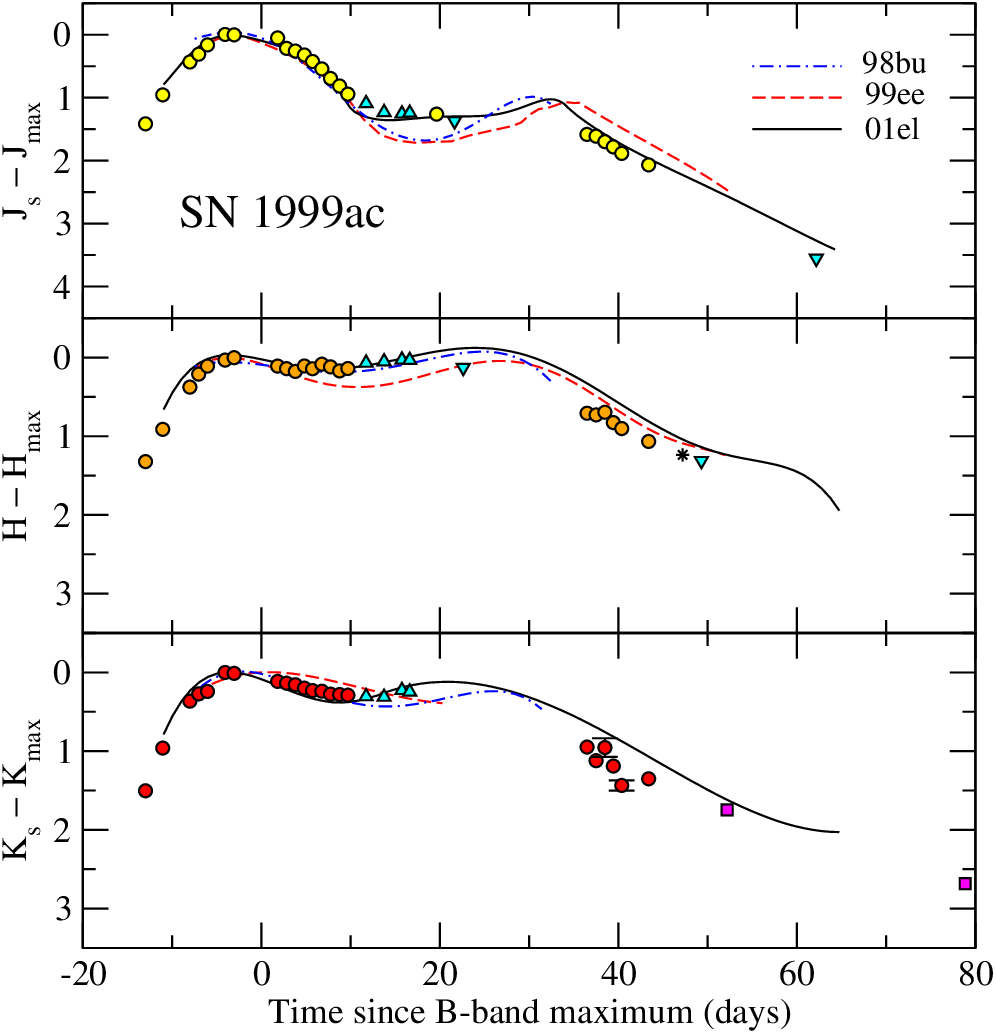]
{$J_s$, $H$, and $K_s$ photometry of SN~1999ac, coded by telescope
and instrument.  For most measurements the uncertainties are less
than or equal to the size of the points.  Symbols:
circles = LCO 1~m; upward pointing triangles = LCO 2.5~m + ClassicCam;
downward pointing triangles = LCO 2.5~m + {\sc cirsi}; asterisk
= Steward Observatory 1.5~m; squares = Steward Observatory 2.3~m.  
For  comparison we have fit higher order polynomials
to the light curves of SNe 1998bu, 1999ee and 2001el.  All data
are plotted with respect to the primary maxima.
The photometry also includes small K-corrections
interpolated from Table 11 of \citet{Kri_etal04b}.  In the time domain
we have subtracted off the dates of $B$-band maximum and corrected for
time dilation.    
\label{jhk}
}

\figcaption[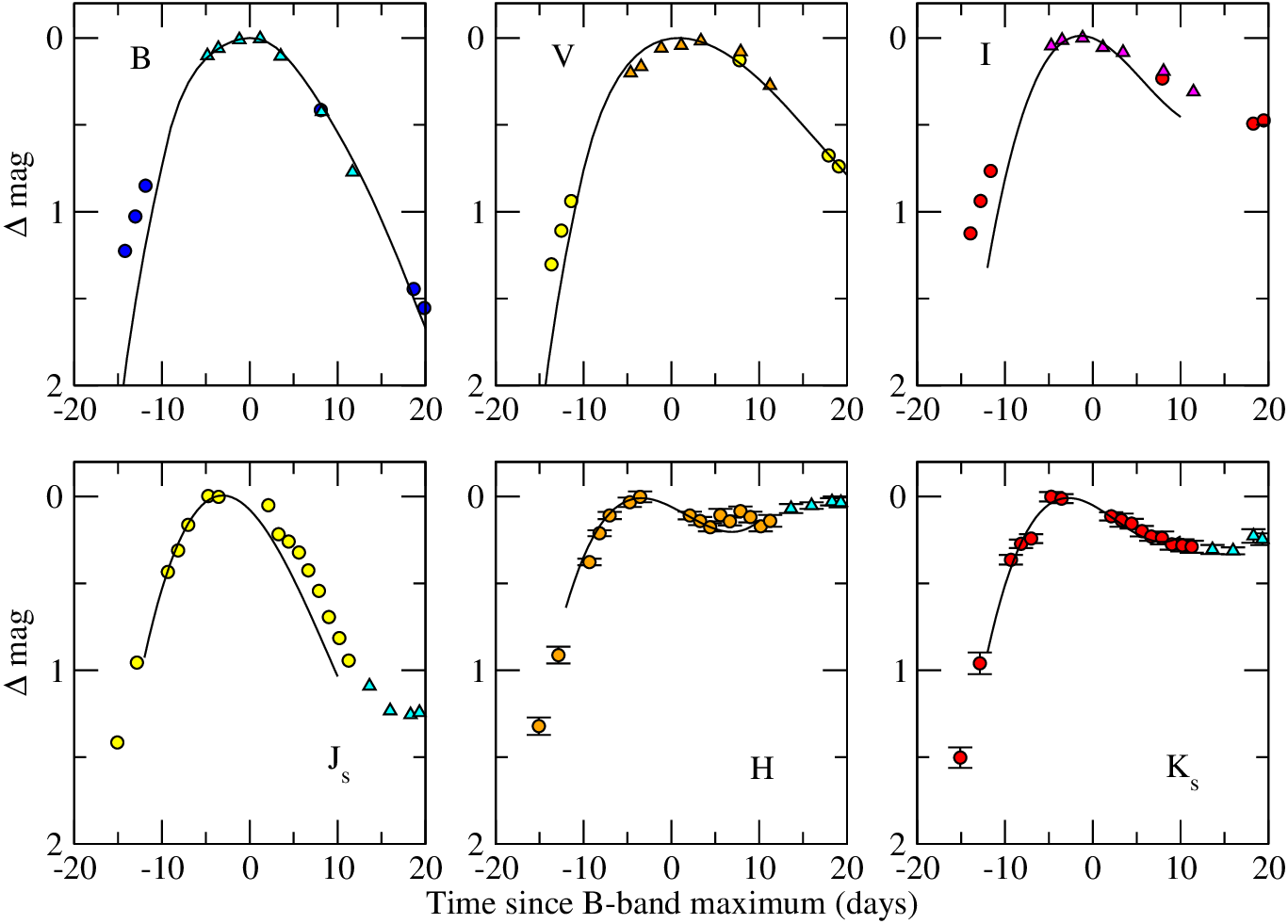]
{Comparison of $BVIJ_sHK_s$ data of SN~1999ac and generic (stretch = 1.00)
light curve templates.  The data have been scaled in the time axis by
stretch factors derived using \dmm\ = 1.33 \citep{Jha02} and also taking into account
time dilation.  So, if the stretched data and
a template match, then the stretch factor correctly characterizes the light
curve.  For the $B$- and $V$-band data the stretch factors are 1.175 and
1.128, respectively.  For the other filters we used the average
of these values, 1.151.  The $B$-band template is from \cite{Gol_etal01},
using a parabolic turn on at $-$20 d.  The $V$-band template is from \citet{Kno_etal03}.
The other maximum-time templates are from \citet{Kri_etal04b}.  In the $BVI$ plots
the circles are CTIO 0.9-m data, while the triangles are data from \citet{Jha02}.
In the $J_sHK_s$ plots the circles are LCO 1-m data, while the triangles are
data from the LCO 2.5-m using ClassicCam.  If error bars are not shown, they are smaller
than or equal to the size of the points.  Since the stretched data do not conform
well to the $BVIJ_s$ templates, the light curves must be characterized by
different stretch factors before and after maximum light.
\label{bvijhk}
}

\figcaption[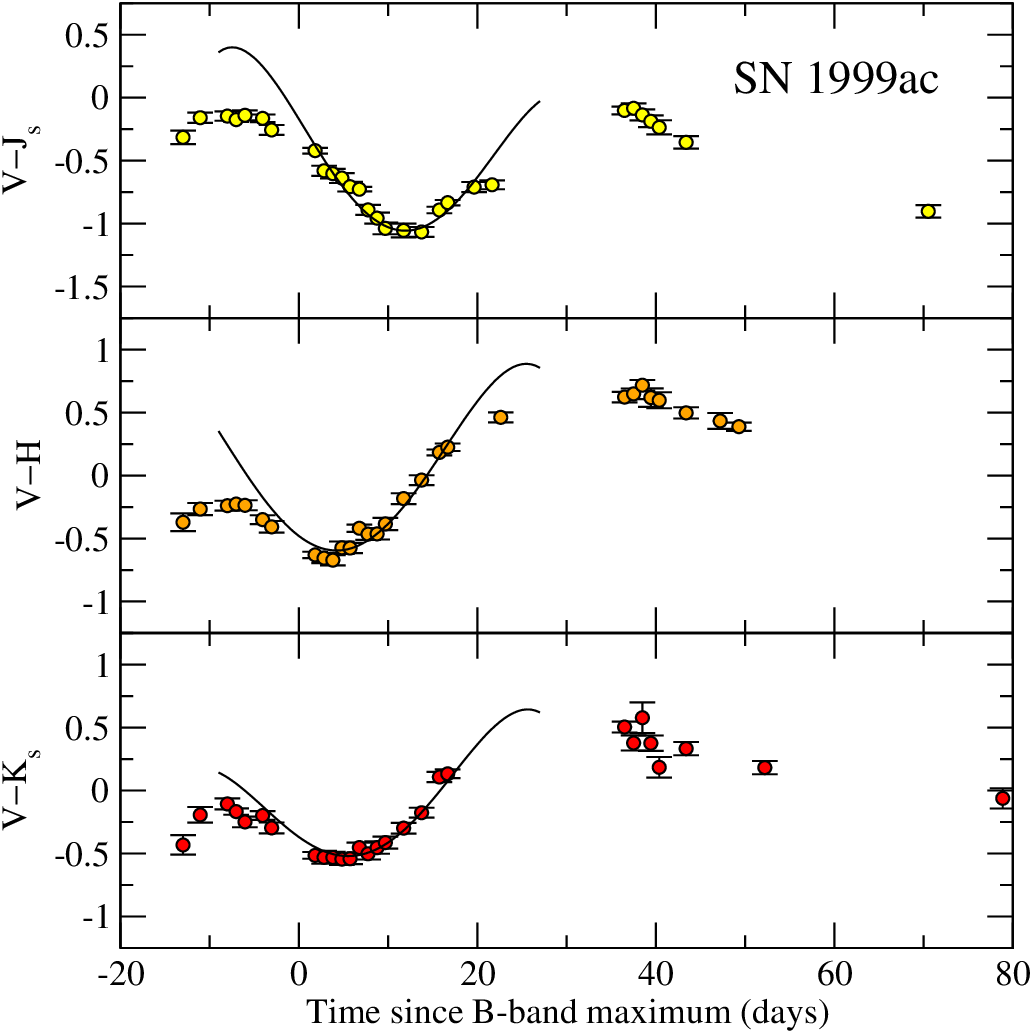]
{$V-J_s$, $V-H$, and $V-K_s$ color evolution of SN~1999ac.  The
solid lines are the zero-reddening loci derived by \citet{Kri_etal00},
based on data for eight SNe~Ia which are mid-range decliners,
and offset by E($V-J$) = 0.44, E($V-H$) = 0.47, and E($V-K$) = 0.37 mag, 
respectively.
\label{vjhk_colors}
}

\figcaption[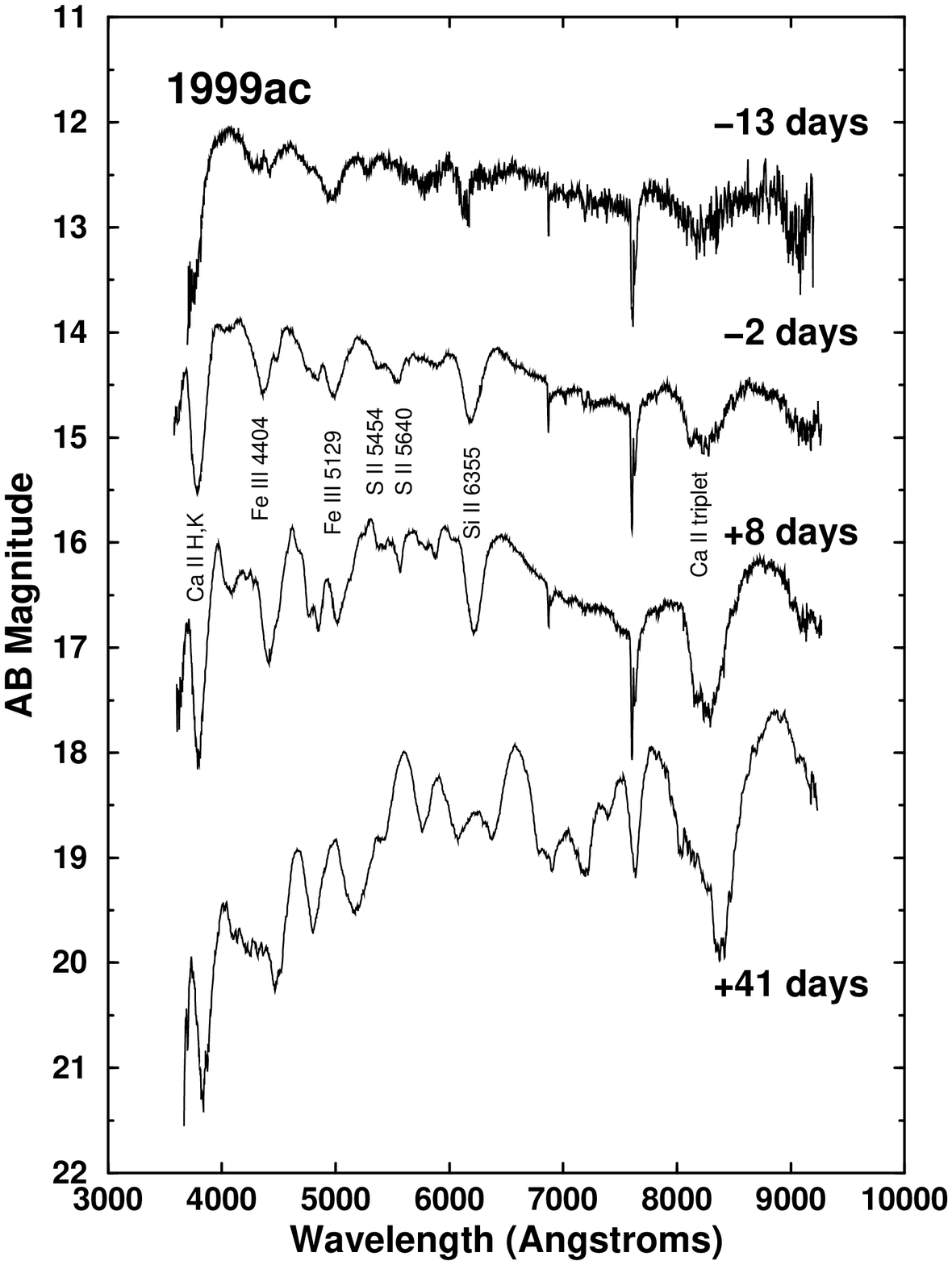]
{Spectra of SN~1999ac obtained at LCO from 13 days before to 
41 days after the epoch of $B$ maximum.  Identifications of the major
contributors to the strongest absorption features in the $-2$ day spectrum
are also indicated.  The abscissa corresponds to observer frame wavelengths.
\label{spectra}
}

\figcaption[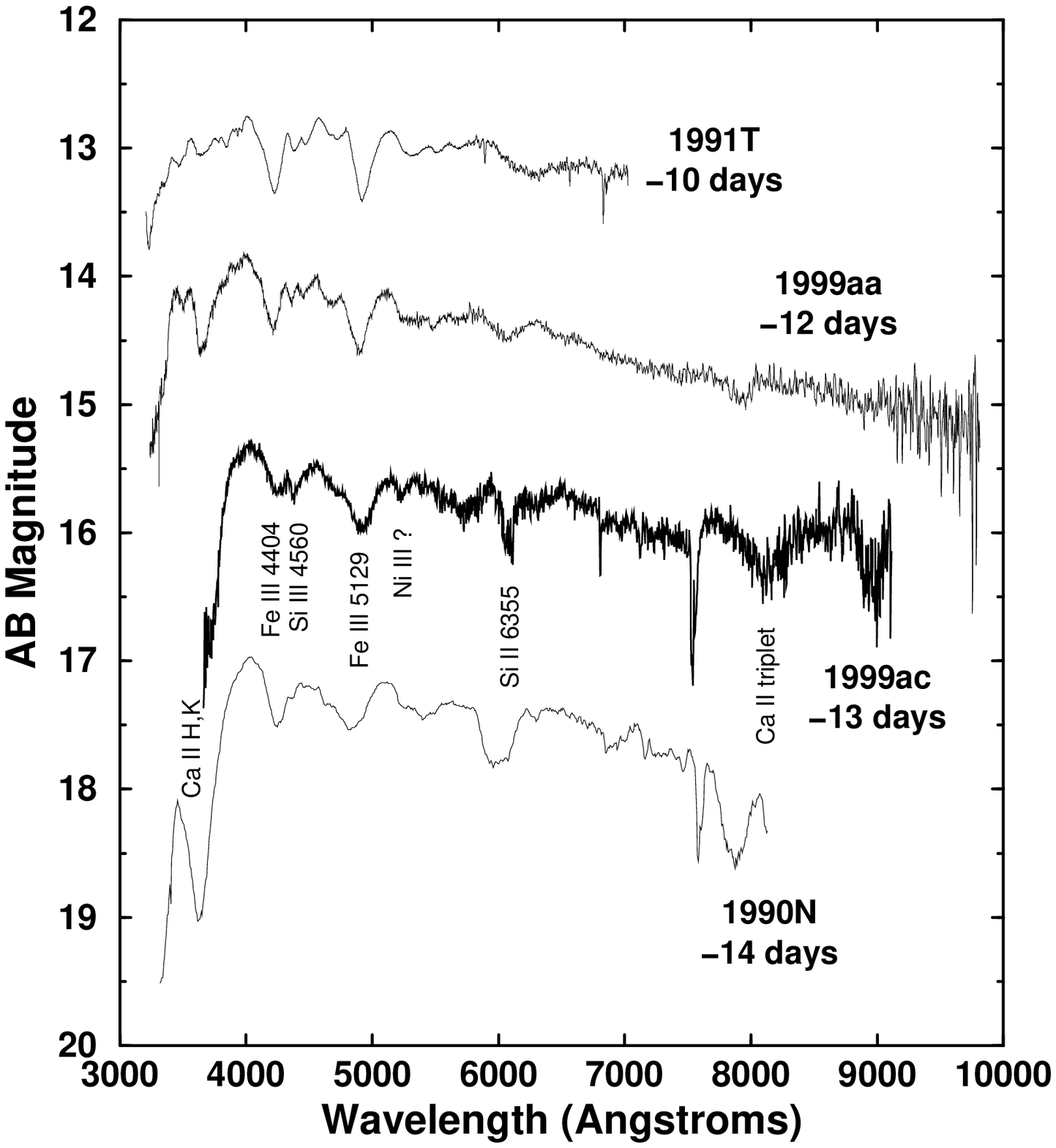]
{Spectrum of SN~1999ac obtained at $-$13 days is compared with
spectra from a similar epoch of SNe~1991T, 1999aa, and 1990N.
Identifications of the major contributors to the strongest absorption
features are indicated.  The abscissa here and in Figs. \ref{comp_m2d}
and \ref{comp_8d} corresponds to rest-frame wavelengths.
\label{comp_m13d}
}

\figcaption[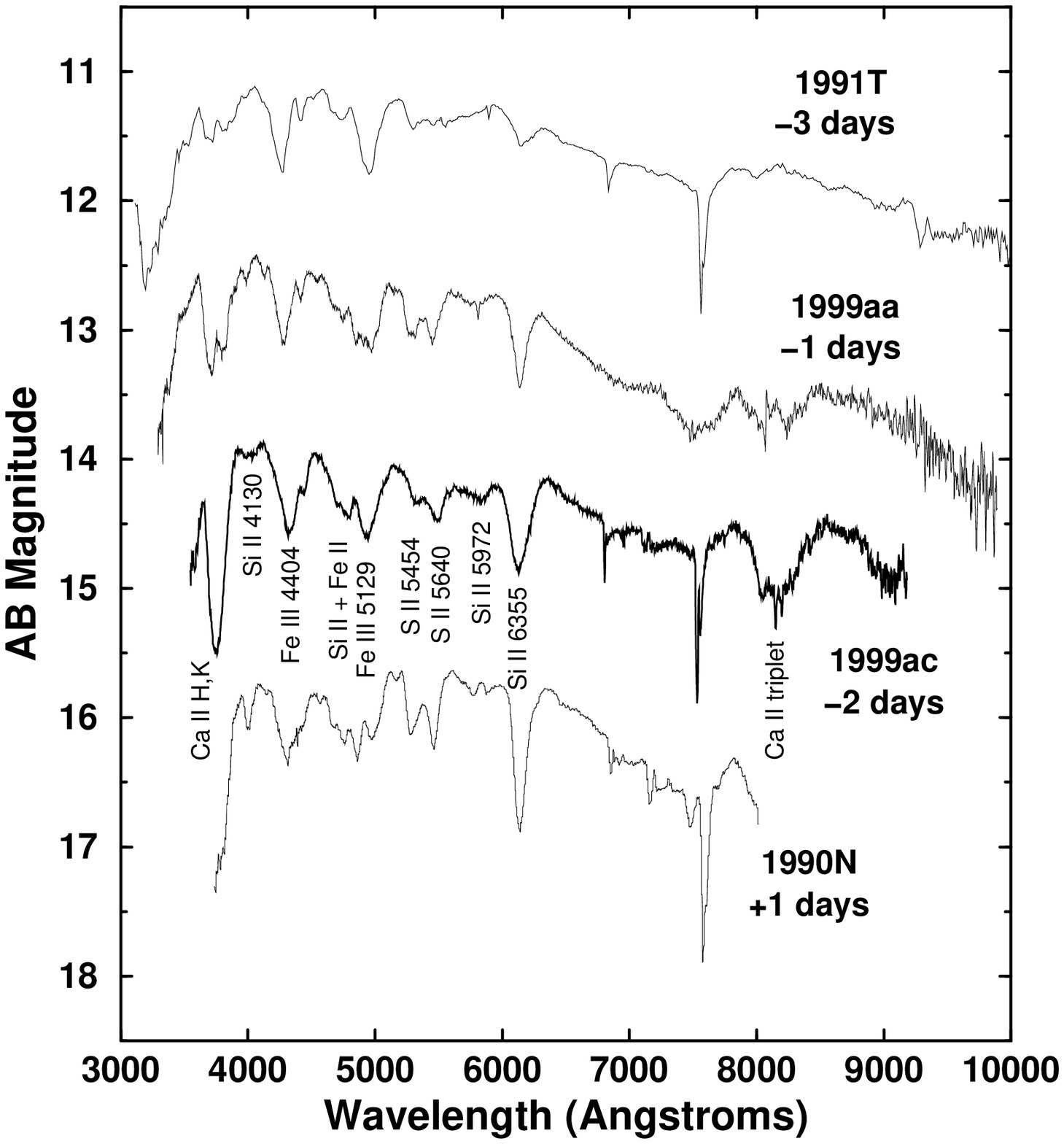]
{Spectrum of SN~1999ac obtained at $-$2 days is compared with
spectra from a similar epoch of SNe~1991T, 1999aa, and 1990N.
Identifications of the major contributors to the strongest absorption
features are indicated.
\label{comp_m2d}
}

\figcaption[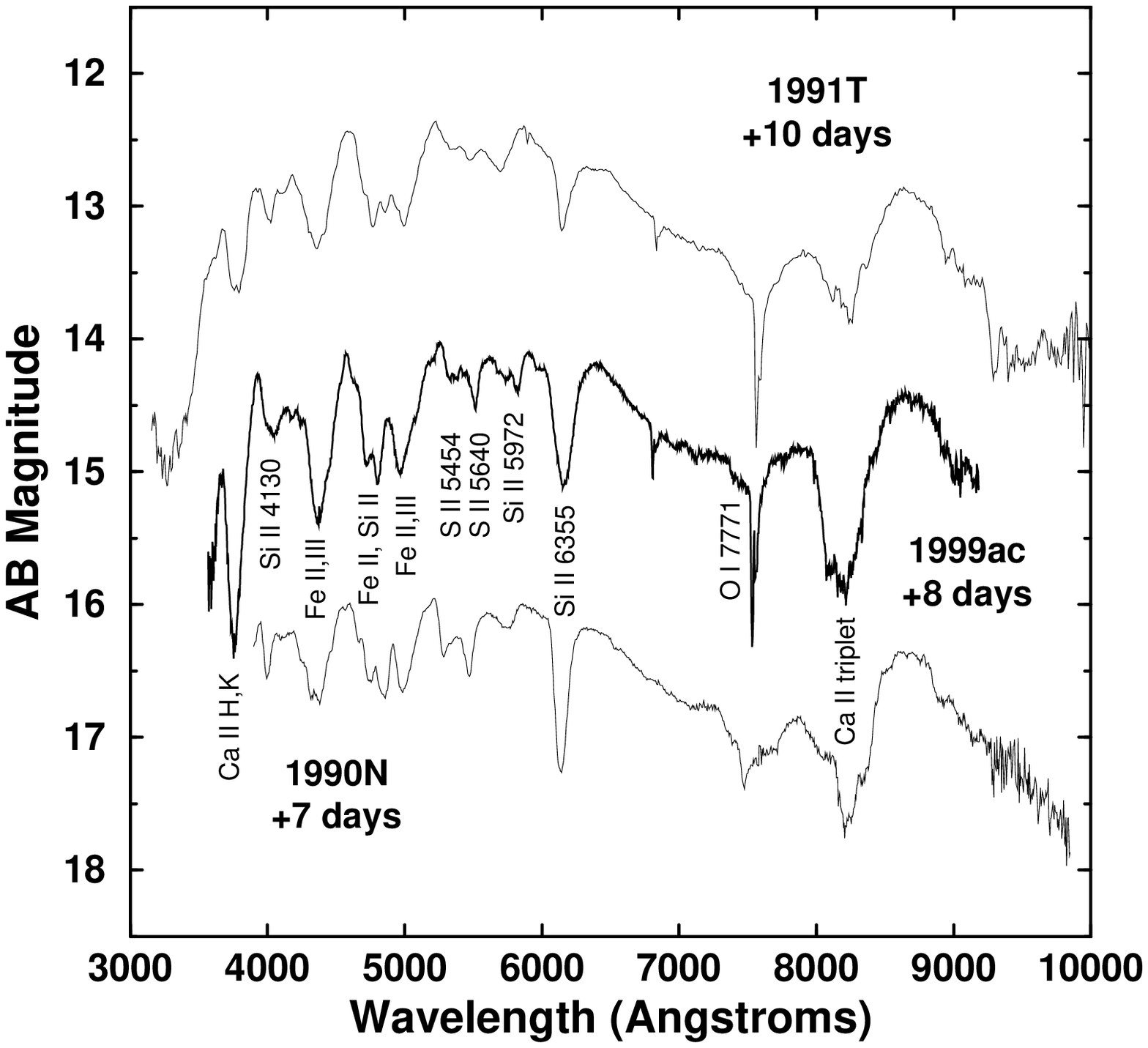]
{Spectrum of SN~1999ac obtained at +8 days is compared with
spectra from a similar epoch of SNe~1991T and 1990N.
Identifications of the major contributors to the strongest absorption
features are indicated.
\label{comp_8d}
}

\figcaption[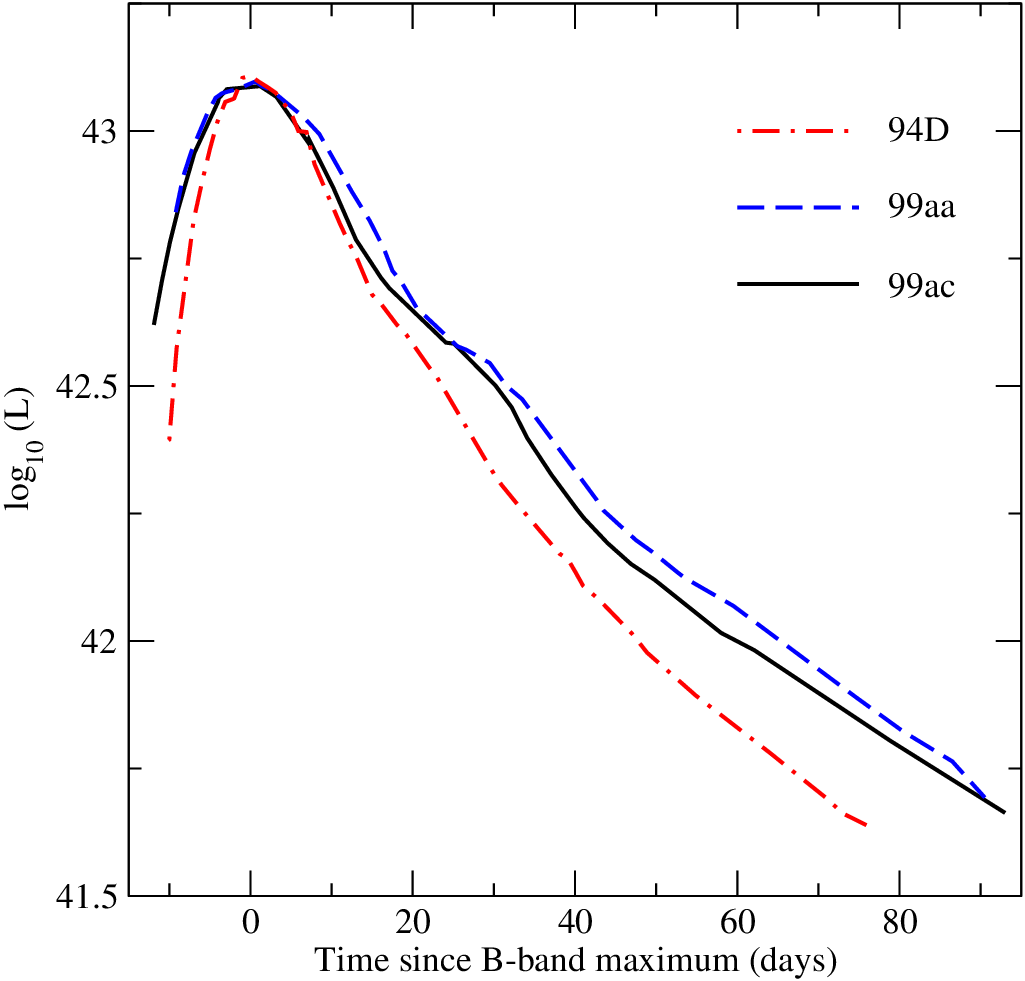]
{Bolometric light curves of SNe~1994D, 1999aa, and 1999ac
constructed from $UBVRI$
photometry.  The luminosity is measured in erg s$^{-1}$.
\label{bol}
}

\figcaption[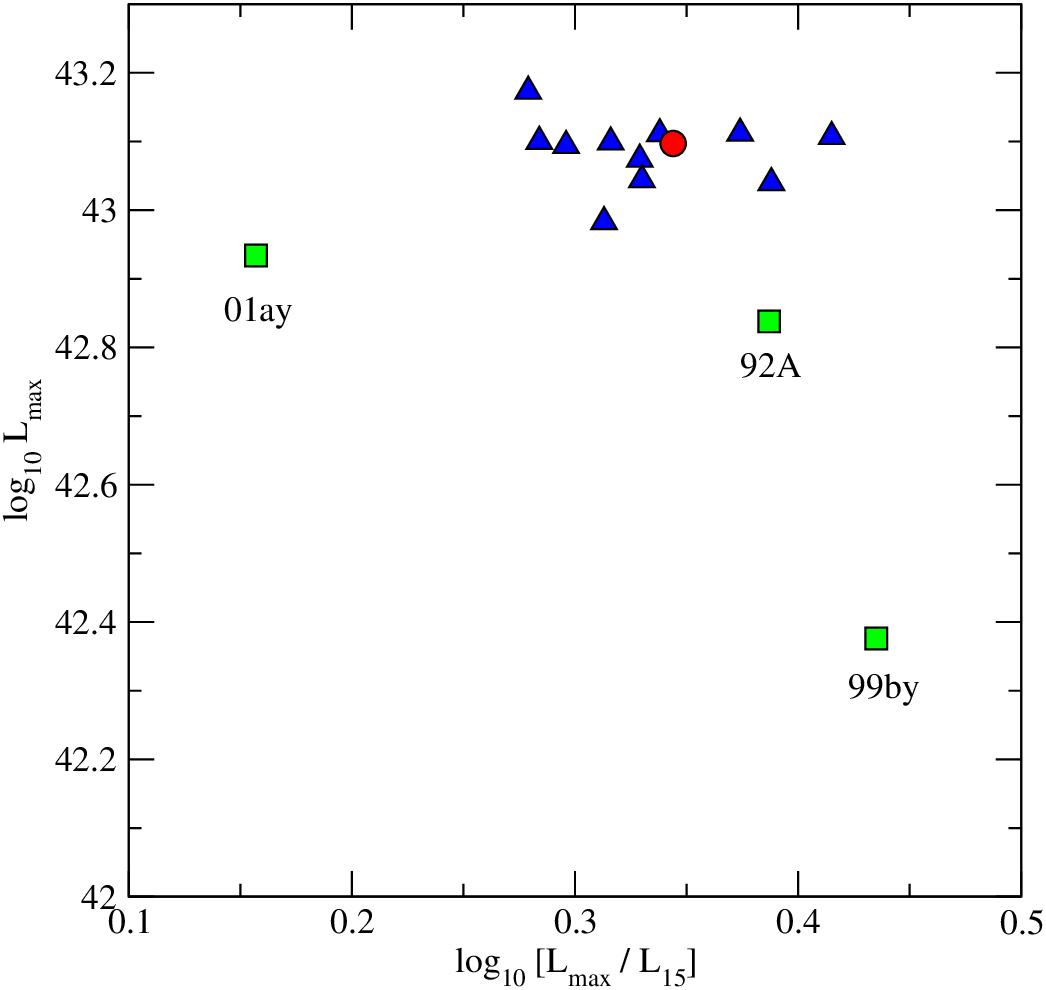]
{Bolometric decline rate relation.  For each object we plot
the logarithm of the peak luminosity (in erg s$^{-1}$) vs.
the bolometric equivalent of \dmm, namely the difference of
log L at the maximum compared to 15 days later.
SN~1999ac is represented
by the (red) circle, while three other specific objects are
labelled.  The remaining objects, represented by triangles,
from left to right are: SNe~1999aw, 1999aa, 1999gp, 1989B,
1990N, 2001el, 1999ee, 1998bu, 1996X, 2000cx, and 1994D.
\label{bol_drr}
}

\clearpage

\begin{figure}
\plotone{sn99ac_finder.ps}
{\center Phillips {\it et al.} Fig. \ref{finder}}
\end{figure}

\begin{figure}
\plotfiddle{bvri.eps}{0.6in}{-90.}{400.}{550.}{-50}{0}
{\center Phillips {\it et al.} Fig. \ref{bvri}}
\end{figure}

\begin{figure}
\plotone{bvri_colors.eps}
{\center Phillips {\it et al.} Fig. \ref{colors}}
\end{figure}

\begin{figure}
\plotone{lcurve_comp.eps}
{\center Phillips {\it et al.} Fig. \ref{bvri_comp}}
\end{figure}

\begin{figure}
\plotone{bvri_colors_fits.eps}
{\center Phillips {\it et al.} Fig. \ref{colors_comp}}
\end{figure}

\begin{figure}
\plotone{sn99ac_jhk.eps}
{\center Phillips {\it et al.} Fig. \ref{jhk}}
\end{figure}

\begin{figure}
\plotone{bvijhk.eps}
{\center Phillips {\it et al.} Fig. \ref{bvijhk}}
\end{figure}

\begin{figure}
\plotone{vjhk_colors.eps}
{\center Phillips {\it et al.} Fig. \ref{vjhk_colors}}
\end{figure}

\begin{figure}
\plotone{spectra.eps}
{\center Phillips {\it et al.} Fig. \ref{spectra}}
\end{figure}

\begin{figure}
\plotone{comp_m13days.eps}
{\center Phillips {\it et al.} Fig. \ref{comp_m13d}}
\end{figure}

\begin{figure}
\plotone{comp_m2days.eps}
{\center Phillips {\it et al.} Fig. \ref{comp_m2d}}
\end{figure}

\begin{figure}
\plotone{comp_8days.eps}
{\center Phillips {\it et al.} Fig. \ref{comp_8d}}
\end{figure}

\begin{figure}
\plotone{bol.eps}
{\center Phillips {\it et al.} Fig. \ref{bol}}
\end{figure}

\begin{figure}
\plotone{bol_drr.eps}
{\center Phillips {\it et al.} Fig. \ref{bol_drr}}
\end{figure}

\end{document}